\documentclass[10pt,a4paper]{article}
\usepackage{jcappub}

\usepackage{color}
\usepackage{amsmath,amssymb}

\newcommand{\kk}{\mathbf{k}}
\newcommand{\pp}{\mathbf{p}}

\newcommand{\qq}{\mathbf{q}}
\newcommand{\xx}{\mathbf{x}}

\newcommand{\vv}{\mathbf{v}}

\newcommand{\LL}{\mathbf{L}}

\newcommand{\YY}{ {\bf \Psi}}

\begin{document}
\title{How does non-linear dynamics affect the baryon acoustic oscillation?}

\author[a,b]{Naonori S. Sugiyama}

\author[b]{and David N.  Spergel}

\affiliation[a]{Astronomical Institute, Graduate school of Science, Tohoku University, Sendai 980-8578, Japan}
\affiliation[b]{Department of Astrophysical Sciences, Peyton Hall, Princeton University, NJ 08544, USA}

\emailAdd{nao.s.sugiyama@gmail.com}

\abstract{

We study the non-linear behavior of the baryon acoustic oscillation in the power spectrum and the correlation function 
by decomposing the dark matter perturbations into the short- and long-wavelength modes. 
The evolution of  the dark matter fluctuations can be described as a 
global coordinate transformation caused by the long-wavelength displacement vector acting on short-wavelength 
matter perturbation undergoing non-linear growth.
Using this feature, 
we investigate the well known cancellation of the high-$k$ solutions in the standard perturbation theory.
While the standard perturbation theory naturally satisfies the cancellation of the high-$k$ solutions,
some of the recently proposed improved perturbation theories do not guarantee the cancellation.
We show that this cancellation clarifies the success of 
the standard perturbation theory at the 2-loop order in describing the amplitude of the non-linear power spectrum even at high-$k$ regions.

We propose an extension of the standard 2-loop level perturbation theory model of the non-linear power spectrum 
that more accurately models the non-linear evolution of the baryon acoustic oscillation than the standard perturbation theory. 
The model consists of
simple and intuitive  parts: the non-linear evolution of the smoothed power spectrum without the baryon acoustic oscillations
and the non-linear evolution of the baryon acoustic oscillations
due to the large-scale velocity of dark matter and due to the gravitational attraction between dark matter particles.
Our extended model predicts the smoothing parameter of the baryon acoustic oscillation peak at $z=0.35$ as $\sim 7.7\ {\rm Mpc}/h$
and describes the small non-linear shift in the peak position due to the galaxy random motions.

}

\keywords{baryon acoustic oscillations, power spectrum}
\arxivnumber{}

\notoc
\maketitle
\flushbottom

\section{Introduction}

Large-scale structure surveys are measuring the galaxy power spectrum
and the position of the baryon acoustic peak with ever increasing precision
\cite{Eisenstein:2005su,Cole:2005sx,Tegmark:2006az,Percival:2006gs,Percival:2009xn,Kazin:2009cj,Beutler:2011hx,
Blake:2010xz,Blake:2011ep,Blake:2011en}.  In the coming decade,  we anticipate that new ground-based surveys such 
as the Prime Focus Spectrograph and Big BOSS and space-based surveys such as Euclid and WFIRST will make even more
accurate measurements of the galaxy power spectrum.  Cosmologists will  use its shape and the position of the BAO peak 
to elucidate the nature of dark energy.

These precision measurements demand precision numerical and theoretical work to describe 
 the non-linear evolution of dark matter and the galaxy power spectrum.  Many different approaches
 have been developed over the past decade for understanding this evolution:
standard perturbation theory 
(SPT; \citep{Bernardeau:2001qr,Fry:1983cj,Goroff:1986ep,Suto:1990wf,Makino:1991rp,Jain:1993jh,Scoccimarro:1996se,Sugiyama:2013,Jeong:2006xd}),
 Lagrangian resummation theory
(LRT; \citep{Matsubara:2007wj,Okamura:2011nu}),
 renormalized perturbation theory
(RPT; \citep{Crocce:2005xy,Crocce:2005xz,Crocce:2007dt}),
 closure theory~\cite{2008ApJ,2009PhRvD},
 multi-point propagator method
(the $\Gamma$-expansion method; \cite{Bernardeau:2008fa,Bernardeau:2011dp,Blas:2013bpa,Bernardeau:2011vy,Bernardeau:2012aq}),
 regularized multi-point propagator method
(Reg PT; \cite{Bernardeau:2011dp,Taruya:2012ut,Taruya:2013my}), 
the Wiener Hermite expansion method \cite{Sugiyama:2012pc},
as well as other techniques~\cite{Pajer:2013jj,Tassev:2012hu,Valageas:2013gba,GilMarin:2012nb,Wang:2012fr,Carlson:2012bu,Tassev:2013pn,Orban:2012uz}.
This plethora of techniques has been both stimulating and at times confusing.  Why do some methods work, while others are less successful at
capturing the non-linear evolution of structure?

In this paper, we explore why the SPT solution succeeds in describing the amplitude of the non-linear power spectrum at high-$k$ regions,
focusing the well known cancellation of its solution at the high-$k$ limit.
We show that the cancellation is related to the fact that the evolution of  the dark matter fluctuations can be described as a 
global coordinate transformation caused by the long-wavelength displacement vector acting on short-wavelength 
matter perturbation undergoing non-linear growth.
While SPT naturally satisfies the cancellation of the high-$k$ solutions,
some of the recently proposed improved perturbation theories, such as RegPT and LRT, do not guarantee the cancellation.
Because of this, we confirm that the SPT solution at the 2-loop order still accurately predicts the non-linear matter power spectrum.

We suggest an improvement to 
the SPT 2-loop solution that more accurately models BAO behavior.
This model has two intuitive parts: 
the smoothed non-linear power spectrum evolution and the non-linear shift of BAO feature.
The non-linear evolution of BAO feature is due to two effects: the primary effect is  due to  the large-scale velocity field and the secondary effect is due to 
 the gravitational correlation effect between dark matter particles.

The $N$-body simulation results used in this paper were presented in \cite{Valageas:2010yw}.
These results and initial conditions at $z_{\rm ini} = 99$ 
were created by the public $N$-body codes {\it GADGET2} and {\it 2LPT} code, respectively~\cite{Springel:2005mi,Crocce:2006ve}.
These $N$-body simulations contain $2048^3$ particles 
and were computed by combining the results with different box sizes $2048 h^{-1}$ ${\rm Mpc}$ and $4096$ $h^{-1}{\rm Mpc}$, called 
$L11$-$N11$ and $L12$-$N11$. 
The cosmological parameters we used were presented by 
the {\it Wilkinson Microwave Anisotropy Probe (WMAP)} five year release~\cite{Komatsu:2008hk}
($\Omega_m = 0.279$, $\Omega_{\Lambda} = 0.721$, $\Omega_b = 0.046$, $h = 0.701$, $n_s = 0.96$ and $\sigma_8=0.817$).
We used the program which is available on Taruya's homepage to compute the predicted power spectra
\footnote{ \url{ http://www-utap.phys.s.u-tokyo.ac.jp/~ataruya/}}.

This paper is organized as follows.
In sections~\ref{Review_of_SPT} and \ref{LPT}, 
we perform the decomposition of the matter perturbation into the long- and short-wavelength terms
in the Eulerian and Lagrangian descriptions, respectively.
In section~\ref{pk},
we show the relation between the long- and short-wavelength decomposition and 
the well known cancellation in the SPT solution at the high-$k$ limit.
Section~\ref{beyond_SPT} presents the non-linear power spectrum model including the more accurate information on BAO than SPT.
In section~\ref{correlation},
we compute the correlation function using our non-linear power spectrum model and investigate the non-linear shift of BAO.
In section~\ref{conclusion}, we summarize and discuss our result.

\section{Standard Perturbation Theory (SPT)}
\label{Review_of_SPT}

\subsection{Review of SPT}

In SPT, the matter density field and the divergence of the velocity density field, $\theta \equiv \nabla \cdot \vv$, is expanded out 
in a perturbation series in the linear growth function, $D$, in Fourier space:
\cite{Bernardeau:2001qr}
\begin{eqnarray}
		\delta(z,\kk) = \sum_{n=1}^{\infty} D^{n} \delta_n(\kk), \quad
		\theta(z,\kk) = -aHf \sum_{n=1}^{\infty} D^{n} \theta_n(\kk),
		\label{expansion}
\end{eqnarray}
where $z$, $H$, and  $a$, are the redshift, the Hubble parameter and  the scale factor.  The velocity
field scales as $f = d\ln D/d\ln a \simeq \Omega^{0.5}$. 

The $n$th-order perturbations are given by
\begin{eqnarray}
		\delta_n(\kk) &=&  \int \frac{d^3p_1}{(2\pi)^3} \cdots \frac{d^3p_n}{(2\pi)^3} 
		(2\pi)^3 \delta_{\rm D}(\kk-\pp_{[1,n]}) F_{n}([\pp_1,\pp_n]) \delta_{\rm lin}(\pp_1) \cdots\delta_{\rm lin}(\pp_n), \nonumber \\
		\theta_n(\kk) &=&   \int \frac{d^3p_1}{(2\pi)^3} \cdots \frac{d^3p_n}{(2\pi)^3} 
		(2\pi)^3 \delta_{\rm D}(\kk-\pp_{[1,n]}) G_{n}([\pp_1,\pp_n]) \delta_{\rm lin}(\pp_1) \cdots\delta_{\rm lin}(\pp_n),
		\label{sol_FG}
\end{eqnarray}
where $F_n(\pp_1,\dots,\pp_n) \equiv F_n([\pp_1,\pp_n])$, $\pp_{[1,n]} \equiv\pp_{1} + \dots + \pp_n $, and
$\delta_1(\kk) \equiv \delta_{\rm lin}(\kk)$.

Under the condition that the amplitudes of $m$ Fourier modes $\pp_{1},\dots,\pp_m$ in $F_n$ and $G_n$
are much smaller than those of the others,
\begin{eqnarray}
		|\pp_{m+1}|, \dots,|\pp_n| \gg |\pp_{1}|, \dots, |\pp_m| \to0,
\end{eqnarray}
the $n$th-order kernel functions are represented by the lower $m$th-order ones than $n$~\cite{Sugiyama:2013}:
\begin{eqnarray}
		F_n(\pp_1,\pp_{n}])\big|_{p_{1},\dots,p_m\to0} &\to& \frac{(n-m)!}{n!}\left(\frac{\pp_{[m+1,n]}\cdot \pp_{1}}{p^2_{1}} \right) 
     	   \cdots \left(\frac{\pp_{[m+1,n]}\cdot \pp_{m}}{p^2_{m}} \right) F_{n-m}([\pp_{m+1},\pp_{n}]), \nonumber \\
		G_n(\pp_1,\pp_{n}])\big|_{p_{1},\dots,p_m\to0} &\to&
		\frac{(n-m)!}{n!}\left(\frac{\pp_{[m+1,n]}\cdot \pp_{1}}{p^2_{1}} \right) 
     	   \cdots \left(\frac{\pp_{[m+1,n]}\cdot \pp_{m}}{p^2_{m}} \right) G_{n-m}([\pp_{m+1},\pp_{n}]). \nonumber \\
		   \label{ap_FG}
\end{eqnarray}

\subsection{Decomposition into the long- and short-wavelength modes in SPT}

What is the physical meaning of the expressions of the kernel functions in eq.~\eqref{ap_FG} ?
To study this, we shall decompose the linear matter density perturbation into the long- and short-wavelength modes as
\begin{eqnarray}
		\delta_{\rm lin}(\kk) = \delta_{\rm lin}^{\rm (L)}(\kk) + \delta_{\rm lin}^{\rm (S)}(\kk),
		\label{decom}
\end{eqnarray}
where the linear long-wavelength mode is defined as $\delta_{\rm lin}(\kk)|_{\kk\to0} \equiv \delta^{(\rm L)}(\kk)$.
We substitute eq.~\eqref{decom} into the solution in SPT [eqs.~\eqref{expansion} and \eqref{sol_FG}]:
\begin{eqnarray}
		\delta(z,\kk) &=&   \sum_{n=1}^{\infty} D^n 
		 \int \frac{d^3p_1}{(2\pi)^3} \cdots \int\frac{d^3p_n}{(2\pi)^3} 
		 (2\pi)^3\delta_{\rm D}(\kk-\pp_{[1,n]})F_{n}([\pp_1,\pp_n])\big|_{p_1,\dots,p_n\to0} 
		\delta_{\rm lin}^{\rm (L)}(\pp_1)\cdots \delta_{\rm lin}^{\rm (L)}(\pp_n) \nonumber \\
		&& + \sum_{n=1}^{\infty} D^n \sum_{m=0}^{n-1}
		\frac{n!}{m!(n-m)!} \int \frac{d^3p_1}{(2\pi)^3} \cdots \int\frac{d^3p_n}{(2\pi)^3} 
		(2\pi)^3\delta_{\rm D}(\kk-\pp_{[1,n]})F_{n}([\pp_1,\pp_n])\big|_{p_1,\dots,p_m\to0} \nonumber \\
		&& \hspace{5cm}
		\times \delta_{\rm lin}^{\rm (L)}(\pp_1)\cdots \delta_{\rm lin}^{\rm (L)}(\pp_m)
		\delta_{\rm lin}^{\rm (S)}(\pp_{m+1}) \cdots \delta_{\rm lin}^{\rm (S)}(\pp_n).
		\label{eq:1}
\end{eqnarray}
The first term in the right-hand side is the contribution from the long-wavelength modes and
the second term is the combination of the long- and short-wavelength modes.
The conditions of $p_1,\dots,p_n \to0$ and $p_{1},\dots,p_m\to0$ are imposed on the kernel functions
because of the long-wavelength modes of the linear matter perturbations.
The first term becomes zero for $n\geq 2$ due to $F_n([\pp_1,\pp_n])|_{p_1,\dots,p_n\to0} (n \geq 2)\to 0$.
This implies that no non-linear effect contributes to the evolution of dark matter in the large-scale limit, an inevitable consequence of mass
and momentum conservation.
Since we are interested
in the small scales where the non-linear effects contribute to the evolution of the matter perturbation,
from now on we ignore the linear matter perturbation in the first term.
Using eq.~\eqref{ap_FG}, we find
\begin{eqnarray}
		\delta(z,\kk)	&=& 
		  \sum_{n=1}^{\infty} D^n \sum_{m=0}^{n-1}
		  \frac{1}{m!} \prod_{i=0}^m\left[ \int \frac{d^3p_i}{(2\pi)^3} 
		  \left( \frac{\pp_{[m+1,n]}\cdot\pp_i}{p_i^2} \right) \delta_{\rm lin}^{\rm (L)}(\pp_i) \right] \nonumber \\
        &&\times
		\int \frac{d^3p_{m+1}}{(2\pi)^3} \cdots \int \frac{d^3p_{n}}{(2\pi)^3}
		(2\pi)^3\delta_{\rm D}(\kk-\pp_{[1,n]}) F_{n-m}([\pp_{m+1},\pp_n]) 
		\delta_{\rm lin}^{\rm (S)}(\pp_{m+1}) \cdots \delta_{\rm lin}^{\rm (S)}(\pp_n) \nonumber \\
		&=&  \sum_{n-m=1}^{\infty} D^{n-m} \int d^3x^{-i\kk\cdot\xx}
       \sum_{m=0}^{\infty}
	   \frac{1}{m!} \left[-i\pp_{[m+1,n]} \cdot \int \frac{d^3p}{(2\pi)^3} e^{i\pp\cdot\xx} D \YY_{\rm lin}^{\rm (L)}(\pp) \right]^m \nonumber \\
		&&\times
		\int \frac{d^3p_{m+1}}{(2\pi)^3} \cdots \int \frac{d^3p_{n}}{(2\pi)^3}
		e^{i\pp_{[m+1,n]}\cdot\xx }F_{n-m}([\pp_{m+1},\pp_n]) \delta_{\rm lin}^{\rm (S)}(\pp_{m+1}) \cdots \delta_{\rm lin}^{\rm (S)}(\pp_n), 
		\label{eq:2}
\end{eqnarray}
where in the second equality we used the summation relation of $\sum_{n=1}^{\infty} \sum_{m=0}^{n-1} = \sum_{n-m=1}^{\infty} \sum_{m=0}^{\infty}$
and $(2\pi)^3 \delta_{\rm D}(\kk-\pp_{[1,n]}) = \int d^3x e^{-i(\kk-\pp_{[1,n]})\cdot\xx}$,
and defined the linear displacement vector as
\begin{equation}
		\frac{i \pp}{p^2} \delta_{\rm lin}(\pp) = \frac{\vv_{\rm lin}(\pp)}{DaH} \equiv {\bf \Psi}_{\rm lin}(\pp).
\end{equation}
Fourier transforming the displacement vector leads to
\begin{eqnarray}
		\delta(z,\kk)&=& \sum_{n=1}^{\infty} D^n \int d^3x e^{-i\kk\cdot\xx}
		\int \frac{d^3p_1}{(2\pi)^3} \cdots\int \frac{d^3p_n}{(2\pi)^3}
		e^{i \pp_{[1,n]}\cdot\left( \xx-D \YY_{\rm lin}^{\rm (L)}(\xx) \right) }
		F_{n}([\pp_{1},\pp_n]) \delta_{\rm lin}^{\rm (S)}(\pp_{1}) \cdots \delta_{\rm lin}^{\rm (S)}(\pp_n). \nonumber \\
		\label{short_k_1}
\end{eqnarray}
In the real space, this expression becomes
\begin{eqnarray}
		\delta(z,\xx) = \delta^{\rm (S)}(z,\xx-D\YY_{\rm lin}^{\rm (L)}(\xx)),
\end{eqnarray}
where the short-wavelength matter perturbation in the real space is defined as
\begin{eqnarray}
		\delta^{\rm (S)}(z,\xx) \equiv 
		\sum_{n=1}^{\infty} D^n
		\int \frac{d^3p_1}{(2\pi)^3} \cdots\int \frac{d^3p_n}{(2\pi)^3}
		e^{i \pp_{[1,n]} \cdot \xx } F_{n}([\pp_{1},\pp_n]) \delta_{\rm lin}^{\rm (S)}(\pp_{1}) \cdots \delta_{\rm lin}^{\rm (S)}(\pp_n).
\end{eqnarray}

For the extreme case where
the long-wavelength displacement vector behaves as a uniform displacement,
$\YY_{\rm lin}^{\rm (L)}(\pp) \to (2\pi)^3\delta_{\rm D}(\pp) \bar{\YY}_{\rm lin}^{\rm (L)}$ where $\bar{\YY}_{\rm lin}^{\rm (L)}$ is constant,
we obtain the following simple expressions in the Fourier and real spaces, respectively:
\begin{eqnarray}
		\delta(z,\kk) &=&    e^{-i\kk\cdot D\bar{\YY}_{\rm lin}^{\rm (L)}} \delta^{\rm (S)}(z,\kk), \nonumber \\
		\delta(z,\xx) &=&  \delta^{\rm (S)}(z,\xx-D\bar{\YY}_{\rm lin}^{\rm (L)}),
		\label{short_k_2}
\end{eqnarray}
where the short-wavelength matter perturbation in the Fourier space is defined as
\begin{eqnarray}
		\hspace{-0.7cm}
		\delta^{\rm (S)}(z,\kk) &\equiv&\sum_{n=1}^{\infty} D^n \int \frac{d^3p_1}{(2\pi)^3} \cdots \int \frac{d^3p_n}{(2\pi)^3}
		(2\pi)^3 \delta_{\rm D}(\kk-\pp_{[1,n]}) F_n([\pp_1,\pp_n]) \delta_{\rm lin}^{\rm (S)}(\pp_1) \cdots \delta_{\rm lin}^{\rm (S)}(\pp_n).
\end{eqnarray}

The same analysis for $\theta$  leads to
\begin{eqnarray}
		\theta(z,\kk) &=&    e^{-i\kk\cdot D\bar{\YY}_{\rm lin}^{\rm (L)}} \theta^{\rm (S)}(z,\kk), \nonumber \\
		\theta(z,\xx) &=&  \theta^{\rm (S)}(z,\xx-D\bar{\YY}_{\rm lin}^{\rm (L)}),
\end{eqnarray}
with
\begin{eqnarray}
		\theta^{\rm (S)}(z,\kk) &\equiv&-aHf\sum_{n=1}^{\infty} D^n \int \frac{d^3p_1}{(2\pi)^3} \cdots \int \frac{d^3p_n}{(2\pi)^3}
		(2\pi)^3 \delta_{\rm D}(\kk-\pp_{[1,n]}) G_n([\pp_1,\pp_n]) \delta_{\rm lin}^{\rm (S)}(\pp_1) \cdots \delta_{\rm lin}^{\rm (S)}(\pp_n).
		\nonumber \\
\end{eqnarray}

These equations describe our basic picture of representing the matter density perturbation and the velocity divergence of dark matter in terms of  
a coordinate transformation due to the long-wavelength displacement vector acting on the short wavelength modes.
In particular, the extreme long-wavelength mode of the displacement vector $\bar{\YY}_{\rm lin}^{\rm (L)}$
globally changes the spatial coordinates throughout the universe  without contributing to the non-linear evolution of the matter perturbations.
This fact implies that we need only consider the non-linear contributions from 
the short-wavelength modes $\delta_n^{\rm (S)}$ in computing  the non-linear evolution of dark matter.

\section{Lagrangian Perturbation Theory}
\label{LPT}
Our  decomposition of the matter perturbation into the long- and short- wavelength modes [eq.~\eqref{short_k_2}]
easily generalizes to  the Lagrangian description.

\subsection{Review of LPT}
In the Lagrangian description, the spatial coordinates are transformed as
\begin{eqnarray}
		\xx \equiv \qq + \YY(z,\qq),
\end{eqnarray}
where $\YY$ is the displacement vector of dark matter.
Conservation of mass implies that  the density perturbation can be  described as a function of the displacement vector:
\begin{eqnarray}
		\delta(z,\kk) &=&  \int d^3q e^{-i\kk\cdot \qq}  \left(  e^{-i\kk\cdot\YY(z,\qq)} -1   \right) \nonumber \\
		&=& \sum_{n=1}^{\infty} \frac{(-i)^n}{n!}\int \frac{d^3k_1}{(2\pi)^3} \cdots \frac{d^3k_n}{(2\pi)^3}
		(2\pi)^3 \delta_{\rm D}(\kk-\kk_{[1,n]})
		\left[ \kk\cdot\YY(z,\kk_1) \right] \cdots 	\left[ \kk\cdot\YY(z,\kk_n) \right].
		\label{psi-delta}
\end{eqnarray}
The displacement vector is perturbatively expanded as (see also appendix~\ref{LPT_Def})
\begin{eqnarray}
		\YY(z,\kk) = \sum_{n=1}^{\infty}D^n \YY_n(\kk),
\end{eqnarray}
where
\begin{eqnarray}
		\YY_{n}(\kk) = \frac{i}{n!}\int \frac{d^3p_1}{(2\pi)^3} \cdots \frac{d^3p_n}{(2\pi)^3}
		(2\pi)^3 \delta_{\rm D}(\kk-\pp_{[1,n]}) \LL_{n}(\pp_1,\dots,\pp_n) \delta_{\rm lin}(\pp_1) \cdots \delta_{\rm lin}(\pp_n).
\end{eqnarray}

\subsection{Decomposition into the long- and short-wavelength modes in LPT}
Now, we decompose the displacement vector into the long- and short-wavelength modes as
\begin{eqnarray}
		\YY(z,\qq) = \bar{\YY}^{\rm (L)}(z) + \YY^{\rm (S)}(z,\qq).
\end{eqnarray}
The long-wavelength displacement vector is defined as $\bar{\YY}^{\rm (L)}(z) \equiv \YY(z,\qq=0)$
whose Fourier-transformation is $(2\pi)^3\delta_{\rm D}(\kk) \bar{\YY}^{\rm (L)}(z)$.
This decomposition allow us to express the matter perturbation as
\begin{eqnarray}
		\delta(z,\kk) &=&  \int d^3q e^{-i\kk\cdot\qq} \left(  e^{-i\kk\cdot\bar{\YY}^{\rm (L)}(z)} -1   \right)
		+ \int d^3q e^{-i\kk\cdot\qq} e^{-i\kk\cdot \bar{\YY}^{\rm (L)}(z)} \left(  e^{-i\kk\cdot\YY^{(S)}(z,\qq)} -1   \right) \nonumber \\
		&=&  (2\pi)^3 \delta_{\rm D}(\kk)\left(  e^{-i\kk\cdot\bar{\YY}^{\rm (L)}(z)} -1   \right)
        + \int d^3q e^{-i\kk\cdot\qq} e^{-i\kk\cdot \bar{\YY}^{\rm (L)}(z)} \left(  e^{-i\kk\cdot\YY^{(S)}(z,\qq)} -1   \right) \nonumber \\
		&=&  e^{-i\kk\cdot \bar{\YY}^{\rm (L)}(z)} \delta^{\rm (S)}(z,\kk),
		\label{Short_g}
\end{eqnarray}
where the first term in the second line becomes zero,
and in the third line the short-wavelength matter perturbation is defined as
\begin{eqnarray}
		\delta^{\rm (S)}(z,\kk) \equiv 
		\int d^3q e^{-i\kk\cdot\qq} \left(  e^{-i\kk\cdot\YY^{\rm (S)}(z,\qq)} -1   \right).
\end{eqnarray}
In the real space, we have
\begin{eqnarray}
		\delta(z,\xx) &=&  \delta^{(S)}(z,\xx-\bar{\YY}^{\rm (L)}(z)) \nonumber \\
		&=&  \delta^{(S)}(z,\qq + \YY^{\rm (S)}(z,\qq) ).
	  \label{main_1}
\end{eqnarray}
This is the first main result of this paper.
This result includes the non-linear long-wavelength displacement vector
and is clearly a generalized version of eq.~\eqref{short_k_2}.
Note that we do not need any dynamics to derive this expression
as long as the matter perturbation is represented by eq.~\eqref{psi-delta} which results from the law of the conservation of mass.

\section{Power spectrum}
\label{pk}

Interestingly, the expression for the matter perturbation in eq.~\eqref{Short_g} naturally explains the high-$k$ solutions in SPT.
The definitions of the correlation terms at 1- and 2-loop level in SPT are summarized in appendix~\ref{SPT_Def}.
\subsection{Cancellation of high-$k$ solutions in SPT}

The high-$k$ solutions in SPT at the 1-loop order is given by
\begin{eqnarray}
		P_{22,\rm high\mathchar`-k}(k)
		&=&  2 \times 2\int \frac{d^3k_1}{(2\pi)^3}\int \frac{d^3k_2}{(2\pi)^3}
		(2\pi)^3 \delta_{\rm D}(\kk-\kk_1-\kk_2) 
		\left[ F_2(\kk_1,\kk_2)\big|_{\rm \kk_1\to0} \right]^2| P_{\rm lin}(k_1) P_{\rm lin}(k_2), \nonumber \\
		&\to&
		 2 \times2 \int \frac{d^3k_1}{(2\pi)^3}\int \frac{d^3k_2}{(2\pi)^3}
		(2\pi)^3 \delta_{\rm D}(\kk-\kk_2) 
		\left[ \frac{1}{2!}\frac{\kk_1\cdot\kk_2}{k_1^2} \right]^2 P_{\rm lin}(k_1) P_{\rm lin}(k_2), \nonumber \\
		&=& \left( \frac{k^2\sigma_{\rm v, lin}^2}{2}  \right) P_{\rm lin}(k), \nonumber \\
		P_{13,\rm high\mathchar`-k}(k)
		&=&  6 P_{\rm lin}(k) \int \frac{d^3p}{(2\pi)^3}F_3(\kk,\pp,-\pp)\big|_{\rm \pp\to0} P_{\rm lin}(p), \nonumber \\
		&\to& 6 P_{\rm lin}(k) \int \frac{d^3p}{(2\pi)^3}	
		\left[ -\frac{1}{3!}\left( \frac{\kk\cdot\pp}{p^2} \right)^2 \right] P_{\rm lin}(p), \nonumber \\
		&=& - \left( \frac{k^2\sigma_{\rm v, lin}^2}{2} \right) P_{\rm lin}(k),
		\label{ap_P1loop}
\end{eqnarray}
where we used Eq.~(\ref{ap_FG}) and 
\begin{eqnarray}
		\sigma_{\rm v,lin}^2 &=& \frac{1}{3\pi^2} \int dp P_{\rm lin}(p).
\end{eqnarray}
These high-$k$ solutions are also obtained using the decomposition of the short- and long wavelength parts in Eq.~(\ref{Short_g}).
From eq.~\eqref{Short_g}, the second- and third-order matter perturbations in SPT are described as
\begin{eqnarray}
		\delta_2(\kk) &=&  \delta_2^{\rm (S)}(\kk) + \left(-i\kk\cdot  \bar{\YY}_{\rm lin}^{\rm (L)}\right)
		\delta_{\rm lin}^{\rm (S)}(\kk), \nonumber \\
		\delta_3(\kk) &=&  \delta_3^{\rm (S)}(\kk) + \left(-i\kk\cdot  \bar{\YY}_{\rm lin}^{\rm (L)}\right) \delta_2^{\rm (S)}(\kk)
		+ \left(-i\kk\cdot  \bar{\YY}_{\rm 2}^{\rm (L)}\right) \delta_{\rm lin}^{\rm (S)}(\kk)
		+ \frac{1}{2} \left(-i\kk\cdot  \bar{\YY}_{\rm lin}^{\rm (L)}\right)^2 \delta_{\rm lin}^{\rm (S)}(\kk).
\end{eqnarray}
Here, the high-$k$ solutions come from the terms including the decorrelation of the short- and long-wavelength parts:
\begin{eqnarray}
		\langle \delta_2(\kk) \delta_2(\kk') \rangle|_{\rm high\mathchar`-k}
		 &=& \left\langle \left(-i\kk\cdot \bar{\YY}_{\rm lin}^{\rm (L)}\right)
		\left( -i\kk'\cdot \bar{\YY}_{\rm lin}^{\rm (L)} \right) \right\rangle
		\left\langle \delta_{\rm lin}^{\rm (S)}(\kk) \delta_{\rm lin}^{\rm (S)}(\kk') \right\rangle \nonumber \\
		&=& (2\pi)^3\delta_{\rm D}(\kk+\kk') \left[ \left( \frac{k^2\sigma_{\rm v, lin}^2}{2}  \right) P_{\rm lin}(k)\right], \nonumber \\
		2 \langle \delta_3(\kk) \delta_{\rm lin}(\kk') \rangle|_{\rm high\mathchar`-k}
		&=& 2 \left\langle \frac{\left( -i\kk\cdot \bar{\YY}_{\rm lin}^{\rm (L)} \right)^2}{2!} \right\rangle
		\left\langle \delta_{\rm lin}^{\rm (S)}(\kk) \delta_{\rm lin}^{\rm (S)}(\kk) \right\rangle \nonumber \\
		&=&  (2\pi)^3 \delta_{\rm D}(\kk+\kk') 
		\left[  - \left( \frac{k^2 \sigma_{\rm v,lin}^2}{2} \right) P_{\rm lin}(k) \right],
		 \label{S_P1loop}
\end{eqnarray}
where since we are interested in the small scale regions where the non-linear effects cannot be ignored,
we can regard that the linear power spectrum is obtained by $\delta^{\rm (S)}_{\rm lin}$.

Similarly to the derivation used in eq.~\eqref{S_P1loop},
the high-$k$ solutions at the 2-loop order in SPT [eq.~\eqref{corrections:2loop}] are given by
\begin{eqnarray}
	P_{33a, \rm high\mathchar`-k}(k) &=& 
	-\frac{1}{2}\left( \frac{k^2 \sigma_{\rm v,lin}^2}{2} \right)P_{13}(k) 
	- \frac{1}{4} \left(\frac{k^2 \sigma_{\rm v,lin}^2}{2} \right)^2 P_{\rm lin}(k),
			\nonumber \\
	P_{33b, \rm high\mathchar`-k}(k)	&=& \left( \frac{k^2 \sigma_{\rm v,lin}^2}{2} \right)P_{22}(k)
	-\frac{1}{2} \left(\frac{k^2 \sigma_{\rm v,lin}^2}{2}\right)^2 P_{\rm lin}(k)
	+ \left( \frac{k^2\sigma_{\rm v, 22}^2}{2} \right) P_{\rm lin}(k), \nonumber \\
	P_{24, \rm high\mathchar`-k}(k) &=&  - \left( \frac{k^2\sigma_{\rm v,lin}^2}{2} \right)P_{22}(k)
	+ \left( \frac{k^2 \sigma_{\rm v,lin}^2}{2}  \right)P_{13}(k)
	+ \left( \frac{k^2\sigma_{\rm v,lin}^2}{2}\right)^2 P_{\rm lin}(k) 
	+ \left(\frac{k^2\sigma_{\rm v,13}^2}{2}\right) P_{\rm lin}(k),\nonumber \\
	P_{15, \rm high\mathchar`-k}(k)  &=&  - \frac{1}{2}\left( \frac{k^2 \sigma_{\rm v,lin}^2}{2} \right)P_{13}(k)
	-\frac{1}{4} \left(\frac{k^2\sigma_{\rm v,lin}^2}{2}\right)^2P_{\rm lin}(k)
	- \left( \frac{k^2\sigma_{\rm v,1\mathchar`-loop}^2}{2} \right) P_{\rm lin}(k),
    \label{ap_P2loop}
\end{eqnarray}
where
\begin{eqnarray}
    			\sigma^2_{\rm v,22} &\equiv&
		\frac{3}{392\cdot32\pi^4} \int \frac{dp_1}{p_1^3} \frac{dp_2}{p_2^3}
	  K(p_1,p_2) P_{\rm lin}(p_1) P_{\rm lin}(p_2), \nonumber \\
        \sigma^2_{\rm v,13} &\equiv&
		\frac{5}{126\cdot32\pi^4} \int \frac{dp_1}{p_1^3} \frac{dp_2}{p_2^3}
	    K(p_1,p_2)P_{\rm lin}(p_1) P_{\rm lin}(p_2), \nonumber \\
      \sigma^2_{\rm v,1\mathchar`-loop} &\equiv&\sigma_{\rm v,22}^2 + \sigma_{\rm v,13}^2, \nonumber \\
	   &=& \frac{167}{3528 \cdot 32\pi^4} \int \frac{dp_1}{p_1^3} \frac{dp_2}{p_2^3}
		K(p_1,p_2) P_{\rm lin}(p_1) P_{\rm lin}(p_2), 
		\label{dispersions}
\end{eqnarray}
with
\begin{eqnarray}
		K(p_1,p_2) =\left( p_1^2-p_2^2 \right)^4\ln \left( \frac{(p_1+p_2)^2}{(p_1-p_2)^2} \right)
		-\frac{4}{3}p_1p_2(3p_1^6 - 11p_2^2p_1^4 - 11 p_1^2 p_2^4 + 3p_2^6).
\end{eqnarray}
In~\cite{Okamura:2011nu},
the expressions of $\sigma_{\rm v,22}^2$ and $\sigma_{\rm v,13}^2$ 
are represented as $\sigma_{\rm v,22}^2 = {\cal A}_{22}/(3\pi^2)$ and $\sigma_{\rm v,13}^2 = 2{\cal A}_{13}/(3\pi^2)$, respectively
(for details see appendix~\ref{LPT_Def}).
However, the expression of the 1-loop velocity dispersion $\sigma_{\rm v,1\mathchar`-loop}^2$ in eq.~\eqref{dispersions} 
is different from eq.~(41) in \cite{Bernardeau:2012ux} which is shown using the eikonal approximation.
Using the expression of the kernel functions in eq.~(\ref{ap_FG}),
we obtain the same expressions as eq.~(\ref{ap_P2loop}) except for lack of the terms including $\sigma^2_{\rm v,13}$ and $\sigma^2_{\rm v,22}$
(see~\cite{Sugiyama:2013}).
This is the result that the expression in eq.~(\ref{Short_g}) is more general than that in eq.~(\ref{short_k_2}).

\begin{figure}[t]
		\hspace{3cm}
				\psfig{figure=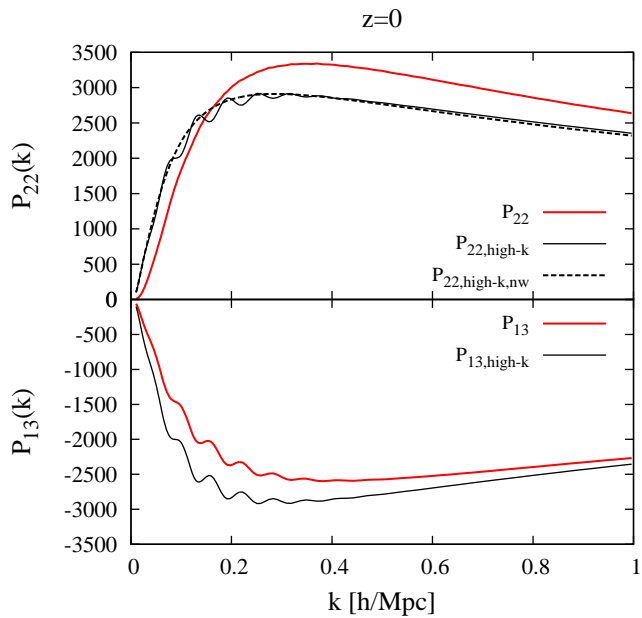}
				\caption{The exact solutions of $P_{22}$ and $P_{13}$ [eq.~\eqref{corrections:1loop}]
				and the high-$k$ solutions without the short-wavelength modes $P_{\rm 22,high-k}$ and $P_{\rm 13,high-k}$ [eq.~\eqref{ap_P1loop}]
				are plotted as the red and black solid lines, respectively.
		The black dashed line denotes the no-wiggle high-$k$ solution for $P_{22}$ [eq.~\eqref{each_ap_1loop_nw}]. }
		\label{fig:SPT1loop}
\end{figure}

\begin{figure}[t]
		\begin{center}
				\psfig{figure=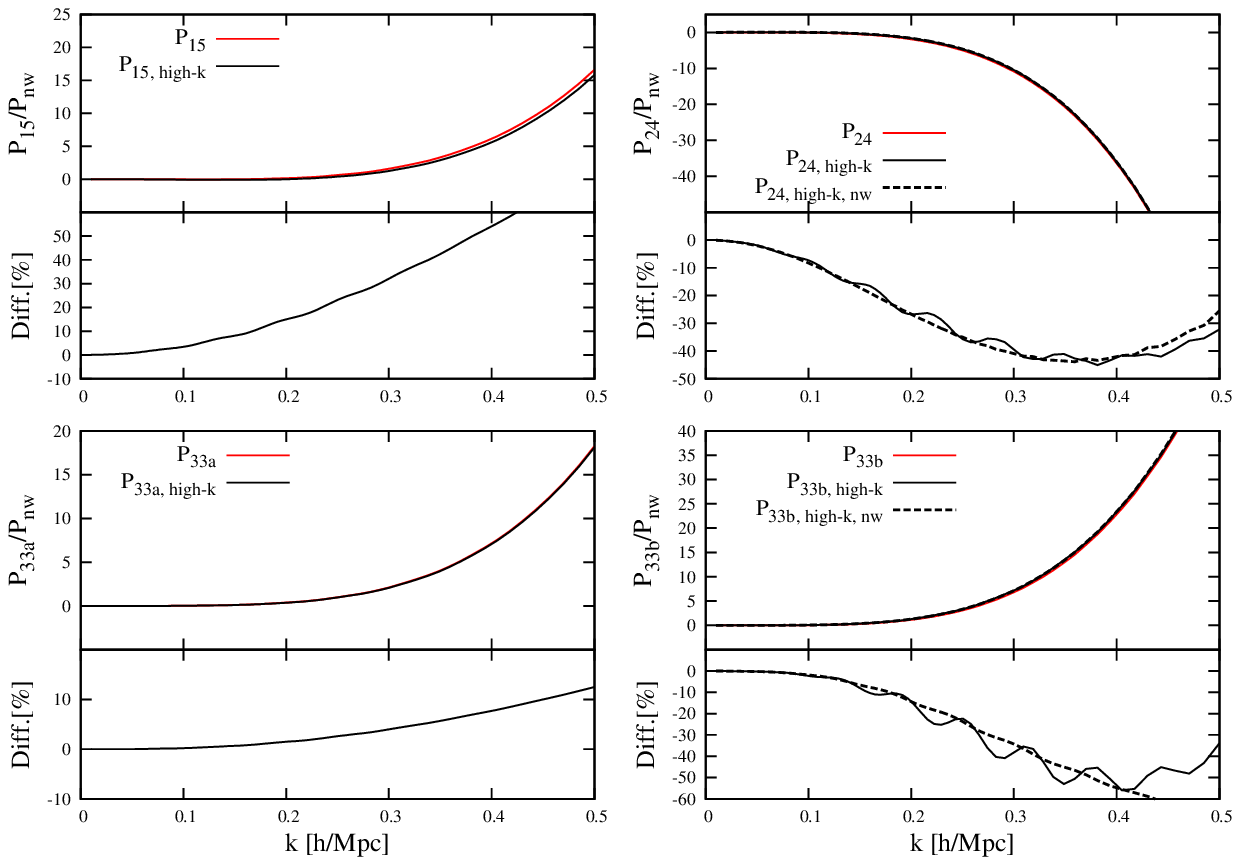}
		\end{center}
		\caption{
		The exact solutions for $P_{15}$, $P_{24}$, $P_{33a}$ and $P_{33b}$ [eq.~\eqref{corrections:2loop}],
		their high-$k$ solutions [eq.~\eqref{ap_P2loop}],
		and the no-wiggle high-$k$ solutions for $P_{24}$ and $P_{33b}$ [eq.~\eqref{each_ap_2loop_nw_1loop}]
		are plotted as the red solid, black solid, and black dashed lines, respectively.
		The fractional differences defined as ${\rm Diff. [\%]}\equiv (P_{\rm exact}- P_{\rm high\mathchar`-k})*100/P_{\rm lin}^{\rm nw}$
		are also plotted as the black solid lines,
		where the no-wiggle linear power spectrum $P_{\rm lin}^{\rm nw}$ is presented in~\cite{Eisenstein:1997ik}.
		For $P_{24}$ and $P_{33b}$,
		the fractional differences between the exact solutions and the no-wiggle high-$k$ solutions are plotted as
		the black dashed lines. }
		\label{fig:SPT2loop}
\end{figure}

Clearly, we find the cancellation of the high-$k$ solutions in SPT in eqs.~(\ref{ap_P1loop}) and (\ref{ap_P2loop}):
\begin{eqnarray}
		P_{\rm 1\mathchar`-loop}(k)|_{\rm high-k} &=&  P_{13,\rm high\mathchar`-k}(k) + P_{22,\rm high\mathchar`-k}(k) = 0, \nonumber \\
		P_{\rm 2\mathchar`-loop}(k)|_{\rm high-k} &=&  P_{33a,\rm high\mathchar`-k}(k) + P_{33b,\rm high\mathchar`-k}(k) 
		                                             + P_{24,\rm high\mathchar`-k}(k) + P_{15,\rm high\mathchar`-k}(k)  = 0.
		\label{Short}
\end{eqnarray}
Figures~\ref{fig:SPT1loop} and \ref{fig:SPT2loop}
show that the high-$k$ solutions have considerable contributions in each solution of SPT even at low-$k$, 
though they completely cancel out each other.
Thus, the cancellation of the high-$k$ solutions in SPT is very important
to accurately compute the non-linear power spectrum even at low-$k$,
and we can understand their origin from the decomposition of the matter density perturbation into the short- and long-wavelength parts.

Here, we define the following quantities which satisfy the cancellation at the high-$k$ limit:
\begin{eqnarray}
		P^{\rm (S)}_{22}(k) &\equiv&  P_{22}(k) - P_{22,\rm high\mathchar`-k}(k), \nonumber \\
		P^{\rm (S)}_{13}(k) &\equiv&  P_{13}(k) - P_{13,\rm high\mathchar`-k}(k),
		\label{OS1}
\end{eqnarray}
and 
\begin{eqnarray}
		P^{\rm (S)}_{33a}(k) &\equiv&  P_{33a}(k) - P_{33a,\rm high\mathchar`-k}(k), \nonumber \\
		P^{\rm (S)}_{33b}(k) &\equiv&  P_{33b}(k) - P_{33b,\rm high\mathchar`-k}(k), \nonumber \\
		P^{\rm (S)}_{24}(k) &\equiv&  P_{24}(k) - P_{24,\rm high\mathchar`-k}(k), \nonumber \\
		P^{\rm (S)}_{15}(k) &\equiv&  P_{15}(k) - P_{15,\rm high\mathchar`-k}(k).
		\label{OS2}
\end{eqnarray}
These quantities yield from the correlation between the long-wavelength displacement vector $\bar{\YY}^{\rm (L)}$ and 
the short-wavelength matter density perturbation $\delta^{\rm (S)}$.
In this paper, we call these quantities the short-wavelength parts of the power spectrum.
Then, the 1- and 2-loop solutions in SPT are
\begin{eqnarray}
		P_{\rm 1\mathchar`-loop}(k) &=&  P_{13}(k) + P_{22}(k) = P_{13}^{\rm (S)}(k) + P_{22}^{\rm (S)}(k)
		= P_{\rm 1\mathchar`-loop}^{\rm (S)}(k), \nonumber \\
		P_{\rm 2\mathchar`-loop}(k) &=&  P_{33a}(k) + P_{33b}(k) + P_{24}(k) + P_{15}(k) 
						  		 =  P_{33a}^{\rm (S)}(k) + P_{33b}^{\rm (S)}(k) + P_{24}^{\rm (S)}(k) + P_{15}^{\rm (S)}(k)
								 =  P_{\rm 2\mathchar`-loop}^{\rm (S)}(k). \nonumber \\
\label{S_2loop}
\end{eqnarray}
The relation between between the original power spectrum and the short-wavelength power spectrum $P^{\rm (S)}$,
such as eqs.~(\ref{OS1}) and (\ref{OS2}), are generalized as follows
\begin{eqnarray}
		P(z,k) 
		= \exp\left( -\frac{\Sigma_{\rm v}^2(z,k)}{2} \right) \exp\left(\frac{\Sigma_{\rm v}^2(z,k)}{2} \right)
		\left[ D^2 P_{\rm lin}(k) + \sum_{n=1}^{\infty} D^{2n+2}P_{\rm n\mathchar`-loop}^{\rm (S)}(k) \right]
		= P^{\rm (S)}(z,k),
		\label{S_power}
\end{eqnarray}
where $\Sigma_{\rm v}^2$ is defined as
\begin{eqnarray}
		\frac{\Sigma_{\rm v}^2(z,k)}{2} \equiv 
		2\sum_{n=1}^{\infty} \frac{(-1)^{n-1}}{(2n)!}  \left\langle \left[ \kk\cdot\bar{\YY}^{\rm (L)}(z) \right]^{2n} \right\rangle_{\rm c},
\end{eqnarray}
and $\langle \cdots \rangle_{\rm c}$ denotes the cumulant.
The definition of $\Sigma_{\rm v}^2/2$ coincides with eq.~(9) in~\cite{Matsubara:2007wj}.
At the 1-loop order, $\Sigma_{\rm v}^2$ is proportional to $k^2$ and becomes
\begin{eqnarray}
		\Sigma_{\rm v}^2(z,k) \equiv k^2 \bar{\Sigma}_{\rm v}^2(z)= k^2D^2\sigma_{\rm v,lin}^2 + k^2D^4 \sigma_{\rm v,1\mathchar`-loop}^2.
		\label{Sigma_tree_1loop}
\end{eqnarray}

\subsection{Comparison with previous works}

\subsubsection{RegPT}

Before we proceed, we shall briefly review the multi-point propagator method ($\Gamma$-expansion method)
\cite{Bernardeau:2008fa,Bernardeau:2011dp,Taruya:2012ut,Sugiyama:2012pc}.
In the $\Gamma$-expansion method, the full non-linear power spectrum is described as
\begin{eqnarray}
		P(z,k) = \sum_{r=1}^{\infty} P_{\rm \Gamma}^{(r)}(z,k),
		\label{Gamma}
\end{eqnarray}
where $P_{\rm \Gamma}^{(r)}$ is the $r$th-order contribution to the power spectrum in the $\Gamma$-expansion, defined as
\begin{equation}
		P_{\rm \Gamma}^{(r)}(z,k) \equiv  r! \int \frac{d^3k_1}{(2\pi)^3} \cdots  \int \frac{d^3k_r}{(2\pi)^3}
		(2\pi)^3 \delta_D(\kk-\kk_{[1,r]}) \left[\Gamma^{(r)}(z,[\kk_1,\kk_r])  \right]^2P_{\rm lin}(k_1) \cdots P_{\rm lin}(k_r).
	\label{power-r}
\end{equation}
The relation between the coefficient of the $\Gamma$-expansion and the kernel function in SPT is given by
\begin{eqnarray}
		\Gamma^{(r)}(z,[\kk_1,\kk_r])
		\equiv 	D^r\Gamma^{(r)}_{\rm tree}([\kk_1,\kk_r]) 
		+ \sum_{n=1}^{\infty} D^{r+2n} \Gamma_{\rm n\mathchar`-loop}^{(r)}([\kk_1,\kk_r]),
\end{eqnarray}
where
\begin{eqnarray}
		&&\Gamma_{\rm tree}^{(r)}([\kk_1,\kk_r])\equiv F_r([\kk_1,\kk_r]), \nonumber \\
		&&\Gamma^{(r)}_{\rm n\mathchar`-loop}([\kk_1,\kk_r]) \nonumber \\
		&&\equiv
       \frac{1}{r!}\frac{(r+2n)!}{2^n n!} 
          \int \frac{d^3p_1}{(2\pi)^3} \cdots  \int \frac{d^3p_n}{(2\pi)^3}
		F_{r+2n}([\kk_1,\kk_r],\pp_1,-\pp_1,\dots,\pp_n,-\pp_n)P_{\rm lin}(p_1)\cdots P_{\rm lin}(p_n). \nonumber \\
		\label{G-r}
\end{eqnarray}
At the 2-loop order in SPT,
the loop correction terms are classified by the $\Gamma$-expansion as
\begin{eqnarray}
		P_{\rm \Gamma}^{(1)}(z,k) &=&  \left[\Gamma^{(1)}(z,k) \right]^2P_{\rm lin}(k) 
		=  D^2 P_{\rm lin}(k) + D^4 P_{13}(k) + D^6 P_{33a}(k) + D^6 P_{15}(k), \nonumber \\
		P_{\rm \Gamma}^{(2)}(z,k) &=&  D^4 P_{22}(k) + D^6 P_{24}(k),  \nonumber \\
		P_{\rm \Gamma}^{(3)}(z,k) &=&  D^6 P_{33b}(k).   
		\label{Gamma_SPT}
\end{eqnarray}
Note that $\Gamma^{(1)}$ is defined as
\begin{eqnarray}
		\langle \delta(z, \kk) \delta_{\rm lin}(z=0,\kk') \rangle 
		\equiv (2\pi)^3\delta_{\rm D}(\kk + \kk') \Gamma^{\rm (1)}(z,k) P_{\rm lin}(k),
		\label{pro}
\end{eqnarray}
and called the ``propagator'' in RPT~\cite{Crocce:2005xy,Crocce:2005xz,Crocce:2007dt}
and RegPT~\cite{Bernardeau:2008fa,Bernardeau:2011dp,Taruya:2012ut,Taruya:2013my}.

In our previous work~\cite{Sugiyama:2013},
we gave an alternative explanation for RegPT and proposed its extended version [eq.~(40) in~\cite{Sugiyama:2013}]
using the expression of the kernel functions in eq.~\eqref{ap_FG}.
Here, we review RegPT in our context focusing the cancellation of the high-$k$ solutions in SPT.
For this purpose, we truncate the short-wavelength correction terms in eq.~\eqref{S_power} as follows
\begin{eqnarray}
		P(z,k) &=&  \exp\left(-\frac{k^2D^2\sigma_{\rm v,lin}^2}{2} \right)
		\sum_{n=0}^{\infty} \frac{1}{n!} \left(\frac{k^2D^2\sigma_{\rm v,lin}^2}{2} \right)^n \nonumber \\
		&& \times
		\Bigg[ D^2P_{\rm lin}(k) + D^4\left( P_{22}^{\rm (S)}(k) + P_{13}^{\rm (S)}(k) \right)
		+ D^6 \left( P_{33a}^{\rm (S)}(k) + P_{33b}^{\rm (S)}(k) + P_{24}^{\rm (S)}(k) + P_{15}^{\rm (S)}(k) \right)  \nonumber \\
		&& \hspace{4cm}
		+  D^8 \left(  P_{44a}^{(S)}(k) + \frac{P_{13}^{\rm (S)}(k)P_{15}^{\rm (S)}(k)}{2P_{\rm lin}(k)} \right) 
		+ D^{10} \frac{\left[P_{15}^{\rm (S)}(k) \right]^2}{4P_{\rm lin}(k)}  \Bigg], 
		\label{RegPT}
\end{eqnarray}
where we ignored the 1-loop velocity dispersion $\bar{\Sigma}_{\rm v}^2(z) = D^2\sigma_{\rm v,lin}^2$ 
and considered the exact 1- and 2-loop corrections as well as the partial 3-loop and 4-loop corrections
for the short-wavelength terms: namely
$D^4P_{\rm 1\mathchar`-loop}$, $D^6P_{\rm 2\mathchar`-loop}$,
$D^8\left( P_{44a}^{\rm (S)} + P_{13}^{\rm (S)} P_{15}^{\rm (S)}/(2P_{\rm lin}) \right)$, and $D^{10} (P_{15}^{\rm (S)})^2/(4P_{\rm lin}) $,
respectively.
$P_{44a}^{\rm (S)}$ [eq.~\eqref{P44a}], which satisfies the cancellation at the high-$k$ limit, is given by
\begin{eqnarray}
		P_{44a}(k) = P_{44a}^{\rm (S)}(k) -\frac{1}{2}\left( \frac{k^2 \sigma_{\rm v,lin}^2}{2} \right)P_{24}(k)
		- \frac{1}{4}\left( \frac{k^2 \sigma_{\rm v,lin}^2}{2} \right)^2P_{22}(k)
		+ \left( \frac{k^2 \sigma_{\rm v,lin}^2}{2} \right)
		\left[\frac{P_{13}(k)}{2 P_{\rm lin}(k)} + \frac{k^2 \sigma_{\rm v,lin}^2}{4}  \right]^2 P_{\rm lin}(k). 
		\nonumber \\
\end{eqnarray}
Then, we can rewrite eq.~\eqref{RegPT} as
\begin{eqnarray}
		P(z,k) &=& \sum_{r=1}^{\infty} P_{\rm \Gamma}^{(r)}(z,k) \nonumber \\
		&=&  \exp\left(-\frac{k^2D^2\sigma_{\rm v,lin}^2}{2} \right)
		\sum_{r=1}^{\infty} \frac{1}{(r-1)!}
		\left(\frac{k^2D^2\sigma_{\rm v,lin}^2}{2} \right)^{r-1}
		D^2 P_{\rm lin}(k) \left[1 + D^2 \frac{P_{13}^{(S)}(k)}{2P_{\rm lin}(k)} +  D^4 \frac{P_{15}^{(S)}(k)}{2P_{\rm lin}(k)}\right]^2 
		\nonumber \\
		&& + \exp\left(-\frac{k^2D^2\sigma_{\rm v,lin}^2}{2} \right)
        \sum_{r=2}^{\infty} \frac{1}{(r-2)!}
		\left(\frac{k^2D^2\sigma_{\rm v,lin}^2}{2} \right)^{r-2}
		\left[ D^4 P_{22}^{(S)}(k) + D^6 P_{24}^{(S)}(k) + D^8P_{44a}^{(S)}(k) \right] \nonumber \\
		&& + \exp\left(-\frac{k^2D^2\sigma_{\rm v,lin}^2}{2} \right)
        \sum_{r=3}^{\infty} \frac{1}{(r-3)!}
		\left(\frac{k^2D^2\sigma_{\rm v,lin}^2}{2} \right)^{r-3}
		\left[ D^6P_{33b}^{(S)}(k) \right],
		\label{Ex_RegPT_2loop}
\end{eqnarray}
This expression coincides with the result of the extended Reg PT [ eq.~(40) in~\cite{Sugiyama:2013} ].
The integer $r$ corresponds to the order of the $\Gamma$-expansion.
When we truncate the order of the $\Gamma$-expansion at the third order,
we reproduce the original Reg PT solution at the 2-loop order (for details, see~\cite{Sugiyama:2013}).
Equations~\eqref{RegPT} and \eqref{Ex_RegPT_2loop} lead to
\begin{eqnarray}
		P(z,k) = D^2 P_{\rm lin}(k) + D^4 P_{\rm 1\mathchar`-loop}(k) + D^6 P_{\rm 2\mathchar`-loop}
		 + D^8\left(  P_{44a}^{(S)}(k) + \frac{P_{13}^{\rm (S)}(k)P_{15}^{\rm (S)}(k)}{2P_{\rm lin}(k)} \right) 
		+ D^{10} \frac{\left[P_{15}^{\rm (S)}(k) \right]^2}{4P_{\rm lin}(k)}. \nonumber \\
\end{eqnarray}
This expression is coincident with eq.~(43) in~\cite{Sugiyama:2013}.

Our previous work~\cite{Sugiyama:2013} pointed out that
the RegPT solution behaves as a part of the SPT solution and its predicted power spectrum amplitude tend to be smaller than SPT
(see figure 2 in~\cite{Sugiyama:2013}).
Now, we can understand this reason.
In RegPT, the higher loop correction terms than the 2-loop order in SPT which satisfy the cancellation at the high-$k$ limit are only three terms:
$D^4P_{\rm 1\mathchar`-loop}$, $D^6P_{\rm 2\mathchar`-loop}$,
$D^8\left( P_{44a}^{\rm (S)} + P_{13}^{\rm (S)} P_{15}^{\rm (S)}/(2P_{\rm lin}) \right)$, and $D^{10} (P_{15}^{\rm (S)})^2/(4P_{\rm lin})$.
However, their correction terms hardly contribute to the power spectrum at large scales (see figure 1 in~\cite{Sugiyama:2013}).
On the other hand, at small scales where they can not be ignored, the exponential damping factor becomes dominant.
Therefore, the amplitude of the solution in RegPT always tends to be smaller than that in SPT.

The propagator is useful for analyzing BAO, described by summing up all terms proportional to the linear power spectrum:
\begin{eqnarray}
		\langle \delta(z,\kk) \delta_{\rm lin}(z=0,\kk') \rangle
		&=&(2\pi)^3 \delta_{\rm D}(\kk+\kk') 
		\left[ 1 + \sum_{n=1}^{\infty} D^{2n} \frac{P_{1(2n+1)}(k)}{2 P_{\rm lin}(k)}  \right] D P_{\rm lin}(k) \nonumber \\
		&=&(2\pi)^3 \delta_{\rm D}(\kk+\kk') \exp\left( -\frac{\Sigma_{\rm v}^2(z,k)}{4} \right)
		\left[ 1 + \sum_{n=1}^{\infty}D^{2n}\frac{P_{1(2n+1)}^{\rm (S)}(k)}{2P_{\rm lin}(k)} \right] D P_{\rm lin}(k),\nonumber \\
		\label{propagator}
\end{eqnarray}
Thus, we obtain the exponential damping behavior in the propagator by defining the short-wavelength term.
In other words, the above relation is the definition of $P_{1(2n+1)}^{\rm (S)}$.
This exponential damping behavior of the propagator is well known (for example, see~\cite{Bernardeau:2012ux}).
Actually, when we truncate the long- and short-wavelength contributions up to the 2-loop level,
we have the similar expression to eq.~(43) in~\cite{Bernardeau:2012ux}.
\begin{eqnarray}
		\Gamma^{\rm (1)}(z,k) = 
		\exp\left( -\frac{k^2D^2 \sigma_{\rm v,lin}^2 + k^2D^4 \sigma_{\rm v,1\mathchar`-loop}^2}{4}\right)
		\left[ 1 + D^2\frac{P_{13}^{\rm (S)}(k)}{2 P_{\rm lin}(k)} + D^4\frac{P_{15}^{\rm (S)}(k)}{2 P_{\rm lin}(k)}\right]D.
\end{eqnarray}
Again, note that our definition of the 1-loop velocity dispersion [eq.~\eqref{dispersions}] 
is different from eq.~(41) in \cite{Bernardeau:2012ux}.

\subsubsection{Comparison with LRT}

The original LRT was proposed as a theory to describe perturbative regime before shell-crossing,
and as mentioned in [cite], LRT was not designed for the  high-$k$ regime where exponential damping factor is important.
Similarly to RegPT, 
LRT has non-linear correction terms that do not guarantee the cancellation of the high-$k$ solutions in SPT.
In particular, their terms that correspond to 3- and more loop orders in SPT do not satisfy the cancellation.
The regime where LRT is applicable is the low-$k$ regime where this cancellation is not important.

\begin{figure}[t]
		\begin{center}
			\psfig{figure=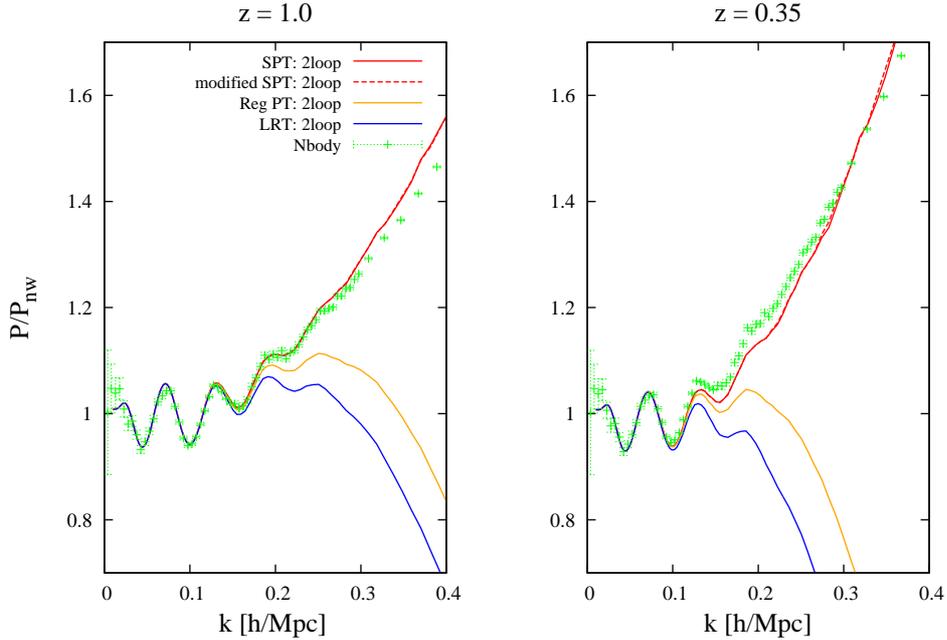}
\end{center}
\caption{The theoretical predictions at the 2-loop level
         (SPT [eq.~\eqref{S_2loop}], modified SPT [eq.~\eqref{main_2}],
		 Reg PT [eq.~\eqref{Ex_RegPT_2loop}], LRT [eq.~\eqref{LRT:2loop}]) and the $N$-body simulation results 
		 are plotted as the red solid, red dashed, orange solid, blue solid lines, and the green symbols
         at $z=1.0$ and $z=0.35$.
		 }
		\label{fig:comparison}
\end{figure}

Equation~\eqref{S_power} coincides
with the generalized formalism of LRT (see eq.~(2.5) in ~\cite{Okamura:2011nu}):
\begin{eqnarray}
		P(z,k) &=&   \exp\left( -\frac{\Sigma_{\rm v}^2(z,k)}{2} \right)
                     \exp\left( \frac{\Sigma_{\rm v}^2(z,k)}{2} \right)
		\left[ D^2P_{\rm lin}(k) + \sum_{n=1}^{\infty}D^{2n+2}P^{\rm (S)}_{\rm n\mathchar`-loop}(k) \right] \nonumber \\
               &=&   \exp\left( -\frac{\Sigma_{\rm v}^2(z,k)}{2} \right)
                     \exp\left( \frac{\Sigma_{\rm v}^2(z,k)}{2} \right)
		\left[ D^2P_{\rm lin}(k) + \sum_{n=1}^{\infty}D^{2n+2}P_{\rm n\mathchar`-loop}(k) \right],
		\label{LRT}
\end{eqnarray}
where we used $P_{\rm n\mathchar`-loop} = P^{\rm (S)}_{\rm n\mathchar`-loop}$.

To derive the LRT solution with the 2-loop corrections,
we truncate the short-wavelength correction terms at the 2-loop level and the velocity dispersion at the 1-loop level,
and partially expand the exponential factor $\exp\left( k^2\bar{\Sigma}_{\rm v}^2/2 \right)$ in eq.~\eqref{LRT} as follows:
\begin{eqnarray}
		P_{\rm LRT,2loop}(z,k)
		&=& \exp\left( -\frac{k^2D^2\sigma_{\rm v,lin}^2 + k^2D^4\sigma_{\rm v,1\mathchar`-loop}^2 }{2} \right)
		\Bigg[  D^2P_{\rm lin}(k) + D^{4}P_{\rm 1\mathchar`-loop}(k)+D^{6}P_{\rm 2\mathchar`-loop}(k) \nonumber\\
		&& 
		+ \left( 1 + 
		\frac{k^2D^2\sigma_{\rm v,lin}^2 + k^2D^4\sigma_{\rm v,1\mathchar`-loop}^2}{2} 
		+ \frac{1}{2!}\left( \frac{k^2D^2\sigma_{\rm v,lin}^2}{2} \right)^2 \right) D^2 P_{\rm lin}(k) \nonumber \\
		&&+ \left( 1 + \frac{k^2D^2\sigma_{\rm v,lin}^2}{2} \right) D^4P_{\rm 1\mathchar`-loop}(k) \Bigg].
		\label{LRT:2loop}
\end{eqnarray}

As shown in figure~\ref{fig:comparison},
RegPT and LRT tend to predict the power spectrum amplitudes that are smaller than SPT
(see also figure 2 in~\cite{Sugiyama:2013})
because of their exponential damping behavior.

\section{Accurate BAO behavior in the power spectrum}
\label{beyond_SPT}

So far, we have showed that 
we need to consider the short-wavelength modes
to get the accurate information on the non-linear evolution of dark matter, 
but not the global coordinate transformation effect from the long-wavelength displacement vector.
However, it is impossible to exactly compute higher order corrections of the short-wavelength modes.
In this section we propose a model for calculating the non-linear power spectrum
including the non-linear shift of BAO in higher order short-wavelength terms than the 2-loop level.

To this end, let us emphasize that 
the first coefficient of the $\Gamma$-expansion $P_{\rm \Gamma}^{(1)} =\left[ \Gamma^{\rm (1)}\right]^2 P_{\rm lin}$
includes the oscillation behavior due to the linear power spectrum,
while by definition the other terms $P_{\rm \Gamma}^{(r)} (r\geq 2)$ have no oscillatory behavior
because they are computed from
the integration of the linear power spectrum and their oscillation behavior in the linear power spectrum is smoothed
[eq.~\eqref{Gamma}] (also see eq.~(9) in~\cite{Crocce:2007dt}).
Furthermore, based on the $\Gamma$-expansion
the full non-linear power spectrum is generally represented as follows: 
\begin{eqnarray}
		P(z,k) = \left[ \Gamma^{(1)}(z,k) \right]^2 P_{\rm lin}^{\rm nw}(k) 
		+ \sum_{n=2}^{\infty} P_{\rm \Gamma}^{(n)}(z,k)
		+ \left[ \Gamma^{(1)}(z,k) \right]^2 \left[ P_{\rm lin}(k) - P_{\rm lin}^{\rm nw}(k) \right],
		\label{rewritten}
\end{eqnarray}
where the no-wiggle linear power spectrum $P_{\rm lin}^{\rm nw}$ is presented in \cite{Eisenstein:1997ik}.
This expression is exact and intuitive.
While the first and second terms represent
the amplitude of the non-linear power spectrum without the oscillation behavior,
the third term only contains the non-linear evolution of BAO.

Remind that we have derived the correction terms in the $\Gamma$-expansion in eq.~\eqref{Ex_RegPT_2loop} for all orders.
However, $P_{\rm \Gamma}^{(n)} (n \geq 2)$ as calculated in eq.~\eqref{Ex_RegPT_2loop} has the oscillatory behavior 
due to the linear power spectrum.
This is because we have truncated the short-wavelength mode at a finite order (the 2-loop level plus some extra terms).
Thus, the lack of the information on higher than 2-loop order short-wavelength terms 
causes the incorrect BAO behavior in the power spectrum.

Therefore,
we propose a simple prescription to modify the incorrect BAO behavior.
We just replace the linear power spectrum in $P_{\rm \Gamma}^{(n)} (n \geq 2)$ calculated in eq.~\eqref{Ex_RegPT_2loop} 
with the no-wiggle linear power spectrum (see also appendix~\ref{high-k}).
Through this replacement, we can get more accurate oscillation behavior in the power spectrum,
even though the amplitude of the power spectrum is not corrected.
Eventually, we get the following expression:
\begin{eqnarray}
		P(z,k) &=&  D^2 P_{\rm lin}^{\rm nw}(k)
		        + D^4 P_{\rm 1\mathchar`-loop}^{\rm nw}(k)
		        + D^6 P_{\rm 2\mathchar`-loop}^{\rm nw}(k) \nonumber \\
				&&
				+ \exp\left( -\frac{k^2\bar{\Sigma}_{\rm v}^2(z)}{2} \right)
				\left[ 1 + D^2\frac{P_{13}^{(\rm S)}(k)}{2P_{\rm lin}(k)} + D^4\frac{P_{15}^{(\rm S)}(k)}{2P_{\rm lin}(k)}  \right]^2
				D^2\left[ P_{\rm lin}(k) - P_{\rm lin}^{\rm nw}(k) \right],
				\label{main_2}
\end{eqnarray}
where the no-wiggle 1- and 2-loop corrections are defined as
\begin{eqnarray}
		P_{\rm 1\mathchar`-loop}^{\rm nw}(k)
		&\equiv& \frac{P_{\rm lin}^{\rm nw}(k)}{P_{\rm lin}(k)} P_{13}(k) + P_{22}(k), \nonumber \\
		P_{\rm 2\mathchar`-loop}^{\rm nw}(k)
		&\equiv& \frac{P_{\rm lin}^{\rm nw}(k)}{P_{\rm lin}(k)} \left[ P_{33a}(k) + P_{15}(k)  \right]
		+ P_{24}(k) + P_{33b}(k),
\end{eqnarray}
and we considered the non-linear velocity dispersion at up to the 1-loop level: 
$\bar{\Sigma}_{\rm v}^2(z) = D^2\sigma_{\rm v,lin}^2 + D^4\sigma_{\rm v,1\mathchar`-loop}^2$
as given in eq.~\eqref{Sigma_tree_1loop}.
This is the second main result of this paper.
The first line in eq.~\eqref{main_2}
corresponds to the first and second terms in eq.~\eqref{rewritten} 
which represent the amplitude of the power spectrum without the BAO behavior.
On the other hand, the second line term in eq.~\eqref{main_2} corresponds to the third term in eq.~\eqref{rewritten}
which only has the non-linear BAO behavior with its amplitude centered around zero.
Note that implicitly in the front of the first line in eq.~\eqref{main_2}
there is the translational symmetry factor $\exp\left( -k^2 \bar{\Sigma}_{\rm v}^2/2 \right)\exp\left( k^2\bar{\Sigma}_{\rm v}^2/2 \right) =1$.
When we expand the exponential factor $\exp\left( -k^2\bar{\Sigma}_{\rm v}^2/2 \right)$ in the second line,
we get the usual SPT 2-loop solution as well as higher order correction terms.
Thus, this model has the more effective information of BAO than the usual SPT 2-loop solution,
and approximately includes higher loop short-wavelength modes than the 2-loop level: $P_{\rm n\mathchar`-loop}^{\rm (S)} (n\geq 3)$.
However, as shown in figure~\ref{fig:comparison},
our modified SPT solution [eq.~\eqref{main_2}] is hardly different from the usual SPT 2-loop solution at BAO scales.
This means that the additional correction terms to the BAO behavior beyond the SPT 2-loop solution [eq.~\eqref{main_2}]
are small enough to be ignored at least at $z=1.0$ and $z=0.35$.

\begin{figure}[t]
		\begin{center}	
				\scalebox{0.8}{\psfig{figure=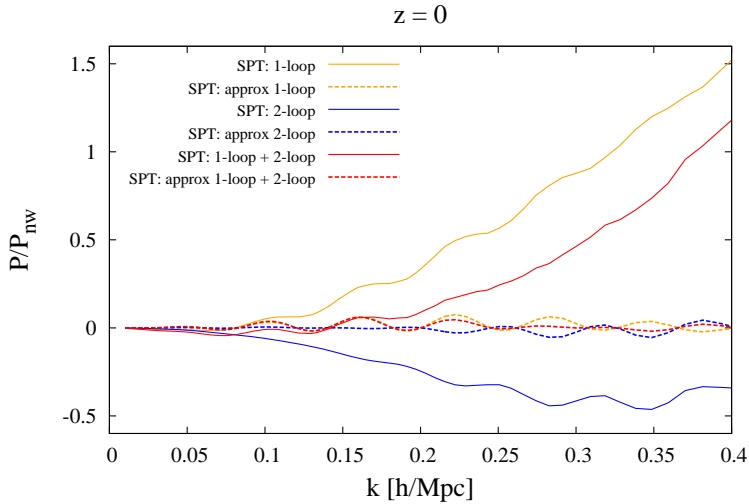}}
		\end{center}	
				\caption{
				The exact an approximated 1- and 2-loop correction terms $P_{\rm 1\mathchar`-loo}$, $P_{\rm 2\mathchar`-loop}$,
				and $P_{\rm 1\mathchar`-loop} + P_{2\mathchar`-loop}$ are plotted 
				as the orange, blue, and red solid (dashed) lines.
				}
		\label{fig:SPT_BAO}
\end{figure}

To test the validity of the above expression [eq.~\eqref{main_2}],
we apply to the same approximation used in the derivation of eq.~\eqref{main_2}
to the linear power spectrum:
\begin{eqnarray}
		P(z,k) &=& \exp\left( -\frac{k^2\bar{\Sigma}_{\rm v}^2(z)}{2} \right)
		        \exp\left( \frac{k^2\bar{\Sigma}_{\rm v}^2(z)}{2} \right)
				D^2P_{\rm lin}(k) \nonumber \\
				&=& \exp\left( -\frac{k^2\bar{\Sigma}_{\rm v}^2(z)}{2} \right) D^2P_{\rm lin}(k)
				+ \exp\left( -\frac{k^2\bar{\Sigma}_{\rm v}^2(z)}{2} \right) 
				  \left( \exp\left( \frac{k^2\bar{\Sigma}_{\rm v}^2(z)}{2} \right) -1 \right) D^2P_{\rm lin}(k) \nonumber \\
				&\to&
        \exp\left( -\frac{k^2\bar{\Sigma}_{\rm v}^2(z)}{2} \right) D^2P_{\rm lin}(k)
				+ \exp\left( -\frac{k^2\bar{\Sigma}_{\rm v}^2(z)}{2} \right) 
				\left( \exp\left( \frac{k^2\bar{\Sigma}_{\rm v}^2(z)}{2} \right) -1 \right) D^2P_{\rm lin}^{\rm nw}(k) \nonumber \\
				&=& 
		D^2P_{\rm lin}^{\rm nw}(k) + \exp\left( -\frac{k^2\bar{\Sigma}_{\rm v}^2(z)}{2} \right)
		D^2\left( P_{\rm lin}(k) - P_{\rm lin}^{\rm nw}(k) \right),
		\label{modified_linear}
\end{eqnarray}
where in the third line we replaced the linear power spectrum of the second term with the no-wiggle linear power spectrum.
This model has no correction to the amplitude of the power spectrum and 
its amplitude is almost the same as the linear power spectrum.
However, this model includes the correction terms to the BAO behavior.
From eq.~\eqref{modified_linear}, the approximated 1- and 2-loop correction terms are described as
(see also appendix~\ref{high-k})
\begin{eqnarray}
		P_{\rm 1\mathchar`-loop}(k) &\to& 
		-\left(\frac{k^2 \sigma_{\rm v,lin}^2}{2}\right)\left( P_{\rm lin}(k) - P_{\rm lin}^{\rm nw}(k) \right), \nonumber  \\
		P_{\rm 2\mathchar`-loop}(k) &\to& \left( \frac{1}{2}\left( \frac{k^2 \sigma_{\rm v,lin}^2}{2} \right)^2
		- \left( \frac{k^2 \sigma_{\rm v,1\mathchar`-loop}^2}{2} \right)\right)\left( P_{\rm lin}(k) - P_{\rm lin}^{\rm nw}(k) \right).
		\label{ap_1_2_loop_nw}
\end{eqnarray}
In figure~\ref{fig:SPT_BAO},
we find that the exact solutions of $P_{\rm 1\mathchar`-loop}$ and $P_{\rm 2\mathchar`-loop}$
have a different phase of BAO and tend to cancel out each other,
yielding the somewhat smoothed correction term [red solid line in figure~\ref{fig:SPT_BAO}].
On the other hand, the approximated 1- and 2-loop solutions [eq.~\eqref{ap_1_2_loop_nw}] 
explain this cancellation of the oscillation behavior [red dashed line in figure~\ref{fig:SPT_BAO}],
even though their amplitudes are centered around zero.
Thus, the modified linear power spectrum indeed has effective corrections to the non-linear BAO behavior.
This fact guarantees that our result [eq.~\eqref{main_2}] includes higher order corrections than the 2-loop level 
and corrects the non-linear evolution of BAO compared to the usual SPT 2-loop solution [eq.~\eqref{S_2loop}].

\begin{figure}[t]
		\begin{center}	
		\psfig{figure=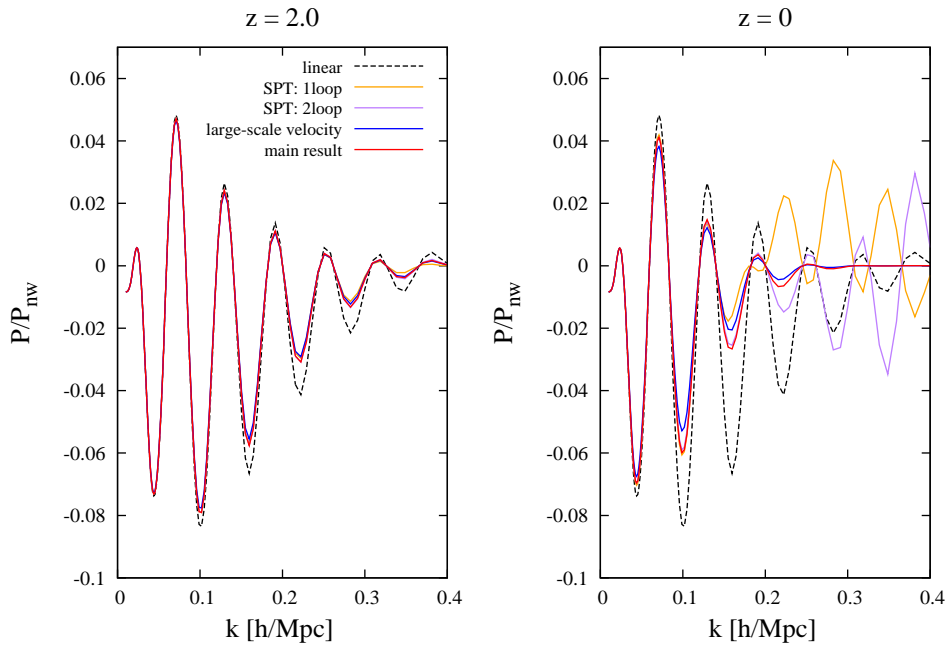}
		\end{center}	
				\caption{The various predictions of the evolution of BAO are plotted.
				The linear, 1-loop, and 2-loop corrections to BAO given in eq.~\eqref{BAO_l12}
				are plotted as the black dashed, orange solid, purple solid lines.
				The BAO behavior in the modified linear model (the second term in eq.~\eqref{modified_linear})
				and our	result (the second line in eq.~\eqref{main_2}) are plotted by the blue and red solid lines, respectively.
				}
		\label{fig:BAO}
\end{figure}

In figure~\ref{fig:BAO},
we compare the various predictions for the non-linear evolution of BAO:
linear, 1-loop, 2-loop, the modified linear model [eq.~\eqref{modified_linear}], and our new
result [eq.~\ref{main_2}].
The linear, 1-loop, and 2-loop BAO behaviors are defined as
\begin{eqnarray}
		P_{\rm BAO, lin}(z,k) &\equiv& D^2\left( P_{\rm lin}(k) - P_{\rm lin}^{\rm nw}(k) \right), \nonumber \\
		P_{\rm BAO, 1\mathchar`-loop}(z,k) &\equiv&
		D^2\left( P_{\rm lin}(k) - P_{\rm lin}^{\rm nw}(k) \right) 
		+ D^4\left( P_{\rm 1\mathchar`-loop}(k) - P_{\rm 1\mathchar`-loop}^{\rm nw}(k) \right) \nonumber \\
		&=& P_{\rm BAO, lin}(z,k) + D^4\left( P_{13}(k) - P_{13}^{\rm nw}(k) \right), \nonumber \\
       	P_{\rm BAO, 2\mathchar`-loop}(z,k) &\equiv& D^2\left( P_{\rm lin}(k) - P_{\rm lin}^{\rm nw}(k) \right)
		+ D^4\left( P_{\rm 1\mathchar`-loop}(k) - P_{\rm 1\mathchar`-loop}^{\rm nw}(k) \right)
		+ D^6\left( P_{\rm 2\mathchar`-loop}(k) - P_{\rm 2\mathchar`-loop}^{\rm nw}(k) \right) \nonumber \\
		&=& P_{\rm BAO, 1\mathchar`-loop}(z,k)
		+ D^6\left(  P_{15}(k) - P_{15}^{\rm nw}(k) + P_{33a}(k) - P_{33a}^{\rm nw}(k) \right).
		\label{BAO_l12}
\end{eqnarray}
At $z=0$ [right panel in figure~\ref{fig:BAO}], 
we find that the SPT 1- and 2-loop solutions have the different BAO behavior from our result, especially at high-$k$.
In particular, the SPT 1-loop solution has a different phase of BAO.
At $z=2.0$, these differences are suppressed.

Let us consider the physical meaning of our result [eq.~\eqref{main_2}].
The exponential factor $\exp\left( -k^2\bar{\Sigma}_{\rm v}^2(z)/2 \right)$ in eq.~\eqref{main_2} comes from the large-scale velocity field 
(the long-wavelength displacement vector $\bar{\YY}^{\rm (L)}$).
This factor causes the acoustic feature to be broader, reducing the amplitude of the acoustic oscillation
(see blue line in figure~\ref{fig:BAO}).
On the other hand, the factor 
\begin{eqnarray}
		\left[ 1 + D^2\frac{P_{13}^{(\rm S)}(k)}{2P_{\rm lin}(k)} + D^4\frac{P_{15}^{(\rm S)}(k)}{2P_{\rm lin}(k)}  \right]^2
		\label{grav}
\end{eqnarray}
is caused by the gravitational effect between dark matter particles.
The gravitational effect restricts the movement of dark matter particles,
preventing the acoustic feature to be spread out by the large-scale velocity of dark matter.
As a result, the reduced amplitude of the BAO peak by the large-scale velocity slightly increases (see red line in figure~\ref{fig:BAO}).

One might wonder why our model [eq.~\eqref{main_2}] has the exponential damping factor computed by the long-wavelength displacement vector.
This is because of
the short-wavelength matter perturbations, but not due to the global coordinate transformation.
The long-wavelength displacement vector can affect the matter power spectrum through the short-wavelength matter perturbation.

We conclude this section by discussing the smoothed linear power spectrum without the BAO feature.
Although the no-wiggle linear power spectrum $P_{\rm lin}^{\rm nw}$ presented in~\cite{Eisenstein:1997ik} has been widely used, 
its amplitude does not exactly coincide with the linear power spectrum, especially in the high-$k$ region.
Therefore, our model [eq.~\eqref{main_2}] largely differs from the usual SPT 2-loop solution [eq.~\eqref{S_2loop}] at small sales.
However, as long as we only focus on BAO scales ($k \leq 0.4 h/{\rm Mpc}$), 
the difference is not important, at least at $z=0.35$ [figure~\ref{fig:comparison}].

\section{Correlation function}
\label{correlation}
\begin{figure}[t]
		\begin{center}
				\psfig{figure=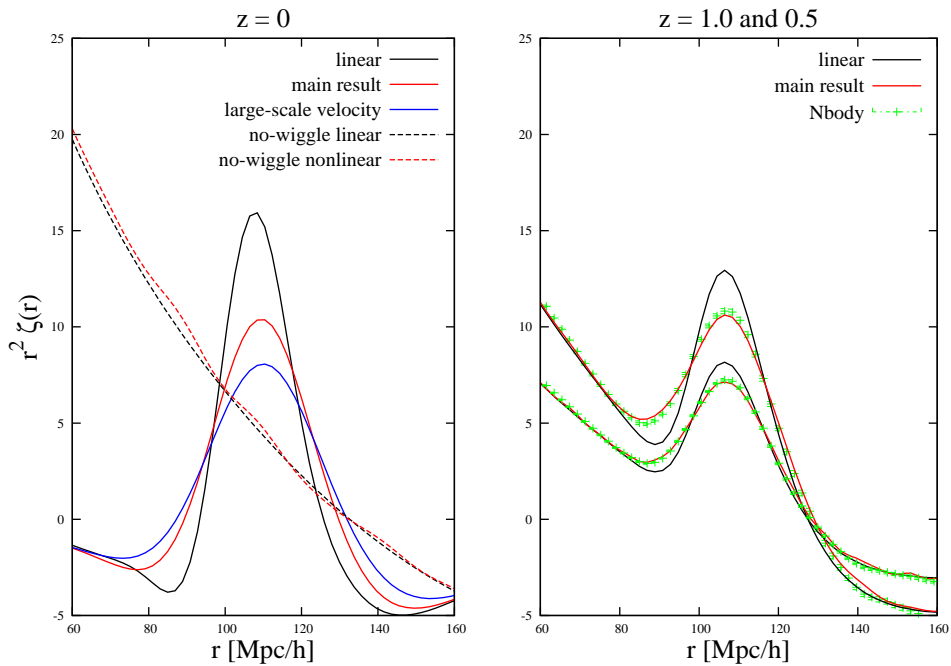}
		\end{center}
				\caption{In the left panel, the smoothed correlation functions without the BAO peak 
				$r^2\xi^{\rm nw}_{\rm lin}$ and $r^2\xi^{\rm nw}_{\rm nl}$ defined
				in eqs.~\eqref{smoothed_corr} 
				and the BAO peak functions $r^2\xi^{\rm BAO}_{\rm lin}$, $r^2\xi^{\rm BAO}_{\rm v}$, and $r^2\xi^{\rm BAO}_{\rm nl}$
				are plotted by the black dashed, red dashed, black solid, blue solid, and red solid lines, respectively at $z=0$.
				In the right panel, the predicted correlation functions computed by
				the linear power spectrum and our model [eq.~\eqref{cut_off_main}],
				and the $N$-body simulations are plotted by the black solid lines, red solid lines, and green symboles
				at $z=1.0$ and $0.5$.
				}
		\label{fig:corr}
\end{figure}

Usually, it is not possible to compute the correlation function in models based on SPT
because of the divergence of the solutions at high-$k$ (see figure.~\ref{fig:SPT_BAO}).
However, truncating the $\Gamma$-expansion (the LRT expansion) at a finite order
leads to exponentially suppressed power spectrum and enables a computation of the correlation function.
In this paper, we also adopt this prescription.
Simply, we use the following expression:
\begin{eqnarray}
		P(z,k) &=& \exp\left( -\frac{k^2\bar{\Sigma}_{\rm v}^2(z)}{2} \right)
		\sum_{n=0}^{6} \frac{1}{n!}\left( \frac{k^2\bar{\Sigma}_{\rm v}^2(z)}{2} \right)^n
				\left[  
                D^2 P_{\rm lin}^{\rm nw}(k)
		        + D^4 P_{\rm 1\mathchar`-loop}^{\rm nw}(k)
		        + D^6 P_{\rm 2\mathchar`-loop}^{\rm nw}(k)  \right]\nonumber \\
				&&
				+ \exp\left( -\frac{k^2\bar{\Sigma}_{\rm v}^2(z)}{2} \right)
				\left[ 1 + D^2\frac{P_{13}^{(\rm S)}(k)}{2P_{\rm lin}(k)} + D^4\frac{P_{15}^{(\rm S)}(k)}{2P_{\rm lin}(k)}  \right]^2
				\left[ D^2P_{\rm lin}(k) - D^2P_{\rm lin}^{\rm nw}(k) \right].
				\label{cut_off_main}
\end{eqnarray}
In the right panel of figure~\ref{fig:corr},
the correlation functions computed by the above model indeed reproduce the $N$-body simulations at $z=1.0$ and $0.5$.
There, we plot $r^2 \xi(r)$ which is proportional to the number of excess pairs in an annulus of width $dr$ centered at $r$. 

To clarify the physical meaning of the non-linear shift of BAO,
we also plot the smoothed correlation functions without BAO 
and the BAO peak functions, defined as (see the left panel in figure~\ref{fig:corr})
\begin{eqnarray}
		\xi_{\rm lin}(z,r) &=&  \xi_{\rm lin}^{\rm nw}(z,r) + 	\xi_{\rm lin}^{\rm BAO}(z,r), \nonumber \\
		\xi_{\rm v}(z,r) &=&  \xi_{\rm lin}^{\rm nw}(z,r) +	\xi_{\rm v}^{\rm BAO}(z,r), \nonumber \\
		\xi_{\rm nl}(z,r) &=&  \xi_{\rm nl}^{\rm nw}(z,r) +	\xi_{\rm nl}^{\rm BAO}(z,r), \nonumber \\
		\label{nw-bao}
\end{eqnarray}
where
\begin{eqnarray}
		\xi_{\rm lin}^{\rm nw}(z,r) &\equiv& \int \frac{dk}{2\pi^2} k^2j_0(kr) D^2P_{\rm lin}^{\rm nw}(k), \nonumber \\
		\xi_{\rm nl}^{\rm nw}(z,r)  &\equiv& \int \frac{dk}{2\pi^2} k^2j_0(kr) 
		\exp\left( -\frac{k^2\bar{\Sigma}_{\rm v}^2(z)}{2} \right)
		\sum_{n=0}^{6} \frac{1}{n!}\left( \frac{k^2\bar{\Sigma}_{\rm v}^2(z)}{2} \right)^n
				\left[  
                D^2 P_{\rm lin}^{\rm nw}(k)
		        + D^4 P_{\rm 1\mathchar`-loop}^{\rm nw}(k)
		        + D^6 P_{\rm 2\mathchar`-loop}^{\rm nw}(k)  \right], \nonumber \\
				\label{smoothed_corr}
\end{eqnarray}
and 
\begin{eqnarray}
		\xi_{\rm lin}^{\rm BAO}(z,r) &\equiv &
		\int \frac{dk}{2\pi^2} k^2j_0(kr) D^2\left[ P_{\rm lin}(k) - P_{\rm lin}^{\rm nw}(k)  \right], \nonumber \\
        \xi_{\rm v}^{\rm BAO}(z,r) &\equiv &
		\int \frac{dk}{2\pi^2} k^2j_0(kr) \exp\left( -\frac{k^2 \bar{\Sigma}_{\rm v}^2(z)}{2} \right)
		D^2\left[ P_{\rm lin}(k) - P_{\rm lin}^{\rm nw}(k)  \right] \nonumber \\
		&=& \int_0^{\infty} ds I(z,r,s) \xi_{\rm lin}^{\rm BAO}(z,s), \nonumber \\
		\xi_{\rm nl}^{\rm BAO}(z,r) &\equiv& \int \frac{dk}{2\pi^2} k^2j_0(kr) 
		\exp\left( -\frac{k^2\bar{\Sigma}_{\rm v}^2(z)}{2} \right)
    	\left[ 1 + D^2\frac{P_{13}^{(\rm S)}(k)}{2P_{\rm lin}(k)} + D^4\frac{P_{15}^{(\rm S)}(k)}{2P_{\rm lin}(k)}  \right]^2
		D^2\left[ P_{\rm lin}(k) - P^{\rm nw}_{\rm lin}(k) \right], \nonumber \\
		\label{BAO_corr}
\end{eqnarray}
where $j_0(x) \equiv \sin(x)/x$, and the function $I$ is
\begin{eqnarray}
		I(z,r,s) \equiv \frac{1}{\sqrt{2\pi \bar{\Sigma}_{\rm v}^2(z)}} \frac{s}{r}
		\left\{  \exp\left[ -\frac{(s-r)^2}{2\bar{\Sigma}_{\rm v}^2(z)} \right]
		-	\exp\left[ -\frac{(s+r)^2}{2\bar{\Sigma}_{\rm v}^2(z)} \right]  \right\}
		\label{smoothed_function}
\end{eqnarray}
and satisfies $\int_0^{\infty} ds I(z,r,s) = 1$.

From eqs.~\eqref{smoothed_corr},
we find that the non-linear evolution of the amplitude of the matter perturbation
hardly affects the correlation function around the BAO peak even at $z=0$ (see the red and black dashed lines in figure~\ref{fig:corr}).
This is because the scales around the BAO peak are too large to neglect the non-linear evolution of the amplitude of 
the matter perturbation.
On the other hand, the dominant contribution to the non-linear shift of BAO comes from $\xi^{\rm BAO}$ in eqs.~\eqref{BAO_corr},
which is computed by the second line in eq.~\eqref{main_2} and the second term in eq.~\eqref{modified_linear}.
As we expected and showed in the case of the power spectrum,
while the large-scale velocity field (the factor $\exp\left( -k^2\bar{\Sigma}_{\rm v}^2/2 \right)$) causes
the BAO peak to be broader,
the gravitational effect between dark matter particles (the factor in eq.~\eqref{grav})
prevents the movement of dark matter particles, slightly increasing the spread BAO peak
(see the blue and red solid lines in the left panel of figure~\ref{fig:corr}).
Thus, the gravitational effect between dark matter particles is a non-neglegible effect
when predicting the non-linear correlation function.

\begin{figure}[t]
		\begin{center}
				\scalebox{0.8}{\psfig{figure=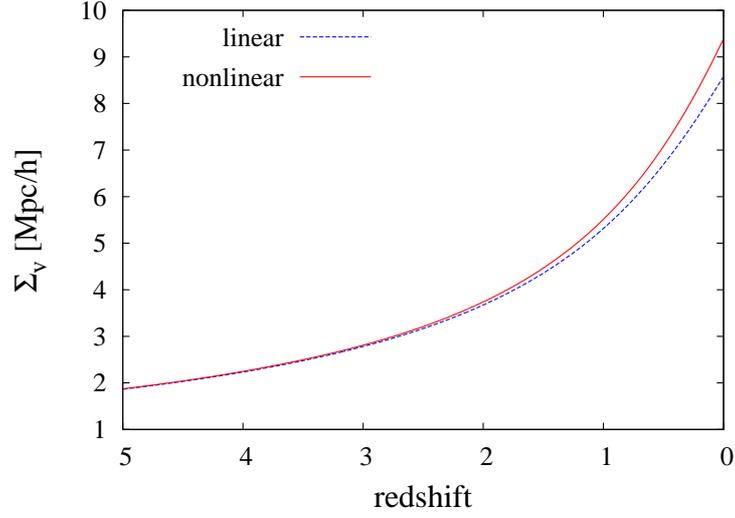}}
		\end{center}
		\caption{Smoothing parameters $\bar{\Sigma}_{\rm v}$ at the linear and 1-loop level
		are plotted as the blue dashed and red solid lines, respectively,
		in the range of redshifts $5 \leq z \leq 0$.}
		\label{fig:sigma}
\end{figure}

We want to mention that the modified linear model [eq.~\eqref{modified_linear}] we presented 
has the same form as the template model to construct the fitting formula 
of the non-linear power spectrum in \cite{Seo:2007ns,Padmanabhan:2012hf,Xu:2012hg,Mehta:2012hh,Xu:2012fw}
\footnote{
See eq.~(14) in \cite{Xu:2012hg},
where the smoothed linear power spectrum and the smoothing parameter 
are represented by $P_{\rm smooth}$ and $\Sigma_{\rm nl}$.
}.
The authors in \cite{Xu:2012hg} fixed the smoothing parameter as $\bar{\Sigma}_{\rm v} = 8\ {\rm Mpc}/h$
before the reconstruction at $z=0.35$.
Now, we can predict the value of $\bar{\Sigma}_{\rm v}$ from eq.~\eqref{Sigma_tree_1loop}.
Figure~\ref{fig:sigma} shows the predicted smoothing parameters in the range of $5.0 \leq z \leq 0$.
At $z=0.35$, we have $\bar{\Sigma}_{\rm v}(z=0.35) = 7.2\ {\rm Mpc}/h$ and $\bar{\Sigma}_{\rm v}(z=0.35) = 7.7\ {\rm Mpc}/h$
at the linear and 1-loop level, respectively.
Thus, we have the correct smoothing parameter by including the 1-loop level velocity dispersion $\sigma_{\rm 1\mathchar`-loop}$.
Furthermore, using our model [eq.~\eqref{main_2}] we could improve the fitting formula of the non-linear power spectrum in \cite{Xu:2012hg}.
This work is left for a future study.

\begin{figure}[t]
		\begin{center}
				\scalebox{1.0}{\psfig{figure=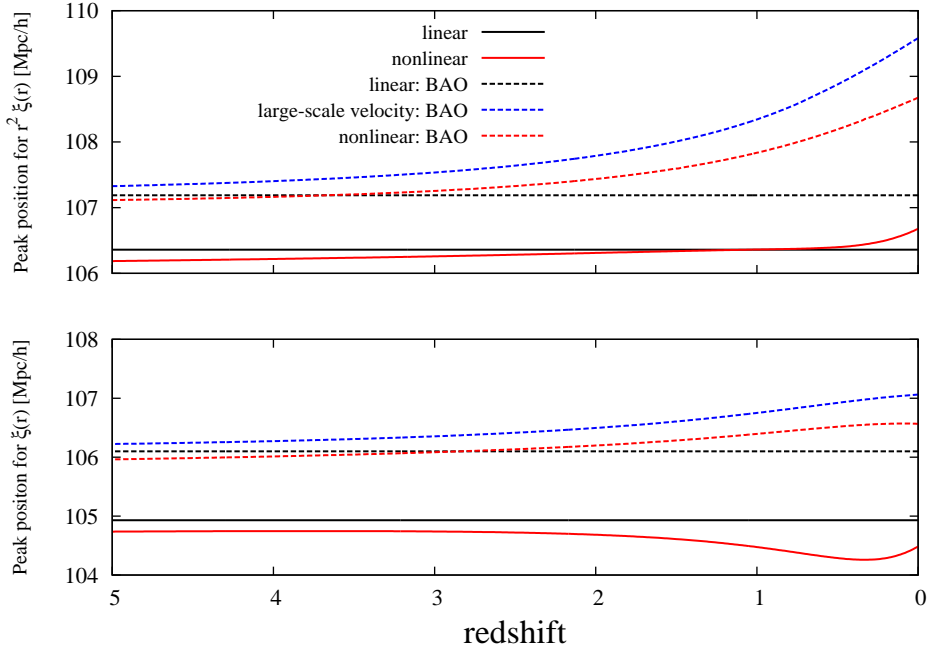}}
		\end{center}
		\caption{Top panel: BAO peak positions of $r^2\xi_{\rm lin}$, $r^2 \xi_{\rm lin}^{\rm BAO}$,
		$r^2\xi_{\rm v}^{\rm BAO}$, $r^2\xi_{\rm nl}^{\rm BAO}$, and $r^2 \xi_{\rm nl}$
		are plotted as the black solid, black dashed, blue dashed, red dashed, and red solid lines, respectively,
		in the range of redshifts $5.0 \leq z \leq 0$.
		Bottom panel: Same as the top panel,
		where the BAO peak positions of 
		$\xi_{\rm lin}$, $ \xi_{\rm lin}^{\rm BAO}$,
		$\xi_{\rm v}^{\rm BAO}$, $\xi_{\rm nl}^{\rm BAO}$, and $ \xi_{\rm nl}$ are plotted.}
		\label{fig:peak}
\end{figure}

\subsection{BAO peak shift}
We consider the shift of the BAO peak position caused by the non-linear evolution.
In Figure~\ref{fig:peak},
we showed the BAO peak positions of
$\xi_{\rm lin}$ (black solid line), $\xi_{\rm lin}^{\rm BAO}$ (black dashed line),
$\xi_{\rm v}^{\rm BAO}$ (red dashed line), $\xi_{\rm nl}^{\rm BAO}$ (red dashed line), and $\xi_{\rm nl}$ (red solid line)
in the range of redshifts $5.0 \leq z \leq 0$.
Their peak positions are numerically computed by the Newton method.

First, we focus on the linear correlation function $r^2 \xi_{\rm lin}$ whose BAO peak is at $\sim 106.4\ {\rm Mpc}/h$.
(see the black solid line in the top panel of Figure~\ref{fig:peak}).
By decomposing into the gravitational collapse effect $r^2 \xi_{\rm lin}^{\rm nw}$ and the BAO effect $r^2 \xi_{\rm lin}^{\rm BAO}$
(eq.~\eqref{nw-bao}),
the linear BAO peak function $r^2 \xi_{\rm lin}^{\rm BAO}$ has its peak at $\sim 107.1\ {\rm Mpc}/h$
(see the black solid line in the left panel of Figure~\ref{fig:corr} and the black dashed line in the top panel of Figure~\ref{fig:peak}).
In other words, the linear gravitational collapse shifts the linear BAO peak position of $r^2 \xi_{\rm lin}^{\rm BAO}$ to small scales.
In the linear theory, this shift of the BAO peak does not depend on redshifts.
Next, the large-scale velocity field effect shifts the BAO peak to large scales.
This feature is expressed by the function $I$ [eq.~\eqref{smoothed_function}] and $\xi_{\rm v}^{\rm BAO}$ [eq.~\eqref{BAO_corr}]
(see the blue line in the left panel of Figure~\ref{fig:corr} and the blue dashed line in the top panel of Figure~\ref{fig:peak}).
We can roughly estimate this BAO peak shift caused by the large-scale velocity field as follows.
Since in the black line in the left panel of Figure~\ref{fig:corr}
the function $r^2 \xi_{\rm lin}^{\rm BAO}$ is very peaked around its peak position $r_{\rm s} \sim 107.1\ {\rm Mpc}/h$,
we assume that $r^2\xi_{\rm lin}^{\rm BAO}$ is proportional to $\delta_{\rm D}(r-r_{\rm s})$.
Then, the spread BAO peak function by the large scale velocity field becomes
\begin{eqnarray}
		r^2\xi_{\rm v}^{\rm BAO}(z,r) &\propto&
		 \frac{1}{\sqrt{2\pi \bar{\Sigma}_{\rm v}^2(z)}} \frac{r}{r_{\rm s}}
		 \left\{  \exp\left[ -\frac{(r-r_{\rm s})^2}{2\bar{\Sigma}_{\rm v}^2(z)} \right]
		 -	\exp\left[ -\frac{(r+r_{\rm s})^2}{2\bar{\Sigma}_{\rm v}^2(z)} \right]  \right\}, \nonumber \\
		 &\sim&  \frac{1}{\sqrt{2\pi \bar{\Sigma}_{\rm v}^2(z)}} \frac{r}{r_{\rm s}}
		\exp\left[ -\frac{(r-r_{\rm s})^2}{2\bar{\Sigma}_{\rm v}^2(z)} \right],
\end{eqnarray}
where the second term in the first line is negligibly small when $r_{\rm s} \gg 1$.
By solving the following equation
\begin{equation}
		\frac{\partial}{\partial r}\left( r^2 \xi_{\rm v}^{\rm BAO}(z,r) \right)
		= 0,
\end{equation}
we have the shifted BAO peak position as
\begin{eqnarray}
		r_{\rm v}^{\rm BAO} &=&  \frac{r_{\rm s}}{2} + \frac{\sqrt{r_{\rm s}^2 + 4\bar{\Sigma}_{\rm v}^2(z)}}{2}, \nonumber \\
		&\sim& r_{\rm s} + \frac{\bar{\Sigma}_{\rm v}^2(z)}{ r_{\rm s}},
\end{eqnarray}
where we used $\bar{\Sigma}^2_{\rm v}/r_{\rm s}^2 \ll 1$.
Thus, the change of the BAO peak position caused by the large-scale velocity field
is estimated using the large-scale velocity dispersion $\bar{\Sigma}_{\rm v}$ as $\bar{\Sigma}_{\rm v}^2/r_{\rm s}$.
Third, the gravitational effect restrict the movement of dark matter particles, slightly increasing the height of the spread BAO peak 
by the large-scale velocity field and shifting the BAO peak position to small scales 
(see the red line in the left panel of Figure~\ref{fig:corr} and the red dashed line in the top panel of Figure~\ref{fig:peak}).
Finally, adding the non-linear gravitational collapse effect $\xi_{\rm nl}^{\rm nw}$ 
the BAO peak position further shifts to small scales
(see the red solid line in the top panel of Figure~\ref{fig:peak}).
Because of the cancellation of the large-scale velocity field effect and the gravitational effect,
the final value of the BAO peak shift becomes small and is less than $1\ {\rm Mpc}/h$.
Also for the correlation function $\xi$,
the same analysis can be done (see the bottom panel in Figure~\ref{fig:peak}).

\section{Conclusion}
\label{conclusion}

We showed that it is possible to
describe the matter perturbation as a global coordinate transformation
from the long-wavelength displacement vector acting on the short-wavelength matter density perturbation.
This feature allows to understand the well known cancellation of the SPT solution at the high-$k$ limit.
The high-$k$ limit contributions in SPT are dominant in each correction term
at the 1- and 2-loop orders, such as $P_{13}$, $P_{22}$, $P_{33a}$, $P_{33b}$, $P_{24}$, and $P_{15}$,
even though their high-$k$ solutions completely cancel out in the total correction terms at the 1- and 2-loop orders.
Some of the recently proposed improved perturbation theories, such as RegPT and LRT,
have correction terms which do not guarantee the cancellation in their expansion approach.
Because of this, they both tend to predict the power spectrum amplitudes that are smaller than SPT.
This implies that SPT still predicts the non-linear power spectrum accurately.

As an example of the model for going beyond SPT,
we presented the model of the non-linear power spectrum
which is based on the SPT 2-loop solution, but more accurately captures the effect of   short-wavelength modes and large scale motions on the BAO peak
position and shape.
From this model,
we have the intuitive behavior of BAO in the correlation function.
We showed that the non-linear evolution of the BAO peak in the correlation function mainly has two effects.
First, the height of the BAO peak is spread by the large-scale velocity field (the long-wavelength displacement vector).
Second, the gravitational effect restricts the movement of dark matter particles, slightly suppressing this spreading of the BAO peak.
The non-linear evolution of the amplitude of the dark matter perturbation does not have an significant influence on  the non-linear BAO evolution.
Our model motivates the template model used to constrain the cosmological parameters from the BAO feature in 
\cite{Padmanabhan:2012hf,Xu:2012hg,Mehta:2012hh,Xu:2012fw},
and correctly predicts the smoothing parameter by including the 1-loop velocity dispersion:
$\bar{\Sigma}_{\rm v} \sim 7.7\ {\rm Mpc}/h$ at $z=0.35$.
Using our model we estimated the value of the BAO peak shift in Figure~\ref{fig:peak},
and showed that the BAO peak shift is less than $1\ {\rm Mpc}/h$ in the range of redshifts $5.0 \leq z \leq 0$.
The reason why the BAO peak shift caused by the non-linear evolution is small
is because of the cancellation between the large-scale velocity field effect and the gravitational effect.
That is, while the large-scale velocity field shifts the BAO peak to large scales,
the gravitational effect shifts it to small scales.
Finally, we note that it might be useful for  BAO peak analyses to use $r^2 \xi - r^2 \xi_{\rm lin}^{\rm nw}$
whose shift of the BAO peak is more visible than $r^2 \xi$,
where we do not need to consider the non-linear gravitational collapse $r^2 \xi_{\rm nl}^{\rm nw}$
because around the BAO peak position the gravitational effect is well represented by the linear theory 
(see the black and red dashed lines in the left panel of Figure~\ref{fig:corr}).

\acknowledgments
We thank T. Futamase, A. Taruya, T. Matsubara, E. Pajer, M. Zaldarriaga, and L. Mercoli for useful comments,
and we thank to T. Nishimiti for providing the numerical simulation results.
This work is supported by a Grant-in-Aid for Scientific Research from JSPS (No. 24-3849 for N.S.S.).
N.S.S. thanks to Departure of Astrophysical Science at Princeton University 
for providing a good environment for research.

\appendix

\section{Definition of the 1- and 2-loop correction terms in SPT}
\label{SPT_Def}

The definitions of the 1- and 2-loop correction terms are given as follows:
\begin{eqnarray}
		P_{\rm 1\mathchar`-loop}(k) &\equiv& P_{22}(k) + P_{13}(k), \nonumber \\
		P_{\rm 2\mathchar`-loop}(k) &\equiv& P_{15}(k) + P_{24}(k) + P_{33a}(k) + P_{33b}(k),
\end{eqnarray}
where
\begin{eqnarray}
		P_{13}(k) &\equiv& 
		6 P_{\rm lin}(k)\int \frac{d^3p}{(2\pi)^3} F_3(\kk,\pp,-\pp) P_{\rm lin}(p), \nonumber \\
		P_{22}(k) &\equiv&  2\int \frac{d^3k_1}{(2\pi)^3}\frac{d^3k_2}{(2\pi)^3} 
		(2\pi)^3 \delta_D(\kk-\kk_{[1,2]}) \left[  F_2(\kk_1,\kk_2)\right]^2 P_{\rm lin}(k_1) P_{\rm lin}(k_2),
		\label{corrections:1loop}
\end{eqnarray}
and 
\begin{eqnarray}
		P_{15}(k)  &\equiv&
		30P_{\rm lin}(k) \int \frac{d^3p_1}{(2\pi)^3}\frac{d^3p_2}{(2\pi)^3}F_5(\kk,\pp_1,-\pp_1,\pp_2,-\pp_2)P_{\rm lin}(p_1)P_{\rm lin}(p_2), 
		\nonumber \\
		P_{33a}(k) &\equiv& \frac{(P_{13}(k))^2}{4P_{\rm lin}(k)}, \nonumber \\
		P_{24}(k)  &\equiv&  24	\int \frac{d^3k_1}{(2\pi)^3}\frac{d^3k_2}{(2\pi)^3}\frac{d^3p}{(2\pi)^3} 
		(2\pi)^3 \delta_{D}(\kk-\kk_{[1,2]}) F_2(\kk_1,\kk_2) F_4(\kk_1,\kk_2,\pp,-\pp)P_{\rm lin}(p) P_{\rm lin}(k_1) P_{\rm lin}(k_2),
		\nonumber \\
		P_{33b}(k) &\equiv& 6	\int \frac{d^3k_1}{(2\pi)^3}\frac{d^3k_2}{(2\pi)^3}\frac{d^3k_3}{(2\pi)^3} 
		(2\pi)^3 \delta_D(\kk-\kk_{[1,3]}) \left[  F_3(\kk_1,\kk_2,\kk_3)\right]^2 P_{\rm lin}(k_1) P_{\rm lin}(k_2) P_{\rm lin}(k_3).
		\label{corrections:2loop}
\end{eqnarray}
Furthermore, a partial correction term at the 3-loop level in SPT $P_{44a}$ is defined as
\begin{eqnarray}
		P_{44a}(k) &\equiv &
		72\int \frac{d^3k_1}{(2\pi)^3}\frac{d^3k_2}{(2\pi)^3}\frac{d^3p_1}{(2\pi)^3}\frac{d^3p_2}{(2\pi)^3} 
		(2\pi)^3 \delta_{\rm D}(\kk-\kk_{[1,2]})\nonumber \\
		&& \times F_{4}(\kk_1,\kk_2,\pp_1,-\pp_1) F_4\left( \kk_1,\kk_2,\pp_2,-\pp_2 \right)
		P_{\rm lin}(k_1)P_{\rm lin}(k_2)P_{\rm lin}(p_1)P_{\rm lin}(p_2).
		\label{P44a}
\end{eqnarray}
\section{Kernel functions of the Lagrangian perturbation theory}
\label{LPT_Def}

We summarize the definition of the kernel function $\LL$ in the Largangian perturbation theory
and the non-linear velocity dispersions $\sigma_{\rm v,1\mathchar`-loop}^2$.

The displacement vector is described as
\begin{eqnarray}
		\YY(z,\kk) = i\sum_{n=1}^{\infty} \frac{D^n}{n!}\int \frac{d^3p_1}{(2\pi)^3} \cdots \frac{d^3p_n}{(2\pi)^3}
		(2\pi)^3 \delta_{\rm D}(\kk-\pp_{[1,n]}) \LL_{n}(\pp_1,\dots,\pp_n) \delta_{\rm lin}(\pp_1) \cdots \delta_{\rm lin}(\pp_n).
\end{eqnarray}
Similarly, the very long-wavelength displacement vector $\bar{\YY}^{\rm (L)}(z)$ is expanded as
\begin{eqnarray}
		\bar{\YY}^{\rm (L)}(z) \equiv i\sum_{n=1}^{\infty}
    \frac{D^n}{n!}\int \frac{d^3p_1}{(2\pi)^3} \cdots \frac{d^3p_n}{(2\pi)^3}
	\LL_{n}(\pp_1,\dots,\pp_n) \delta_{\rm lin}(\pp_1) \cdots \delta_{\rm lin}(\pp_n).
\end{eqnarray}
The kernel functions at up to the third order are given by~\cite{Rampf:2012up}
\begin{eqnarray}
		\LL_1(\pp) &=& \frac{\pp}{p^2}, \nonumber \\
		\LL_2(\pp_1,\pp_2) &=& \frac{3}{7} \frac{\pp_{[1,2]}}{|\pp_{[1,2]}|^2} \left(  1 - \frac{(\pp_1\cdot\pp_2)^2}{p_1^2p_2^2} \right),
		\nonumber \\
		\LL_3(\pp_1,\pp_2,\pp_3) 
		&=& \Bigg\{\frac{5}{21} \frac{\pp_{[1,3]}}{|\pp_{[1,3]}|^2}
		\left( 1 - \frac{(\pp_1\cdot\pp_2)^2}{p_1^2p_2^2} \right)\left( 1 - \frac{(\pp_3\cdot(\pp_1+\pp_2))^2}{p_3^2|\pp_1+\pp_2|^2} \right)
		+\mbox{2 perms.} \Bigg\} \nonumber \\
		&& -\frac{1}{3} \frac{\pp_{[1,3]}}{|\pp_{[1,3]}|^2}
		\left( 1 - \frac{\left( \pp_1\cdot\pp_2 \right)^2}{p_1^2p_2^2}
		-\frac{\left(\pp_1\cdot\pp_3 \right)^2}{p_1^2p_3^2}-\frac{\left( \pp_2\cdot\pp_3 \right)^2}{p_2^2p_3^2} 
		+ 2 \frac{(\pp_1\cdot\pp_2)(\pp_2\cdot\pp_3)(\pp_1\cdot\pp_3)}{p_1^2p_2^2p_3^2}\right) \nonumber \\
		&& + \Bigg\{\frac{1}{7} \frac{\pp_{[1,2]}(\pp_{[1,3]}\cdot\pp_3) - \pp_3( \pp_{[1,3]}\cdot\pp_{[1,2]})}{|\pp_{[1,3]}|^2}
		\left( \frac{\pp_3\cdot\pp_{[1,2]}}{p_3^2|\pp_{[1,2]}|^2}  \right)
		\left( 1 -\frac{(\pp_1\cdot\pp_2)^2}{p_1^2p_2^2} \right)+ \mbox{2 perms} \Bigg\}.
		\nonumber \\
\end{eqnarray}

The non-linear velocity dispersions $\sigma_{\rm v,22}^2$ and $\sigma_{\rm v,13}^2$ are calculated as
\begin{eqnarray}
		k^2\sigma_{\rm v,22}^2 &\equiv&
	- \int \frac{d^3p_1}{(2\pi)^3}\int \frac{d^3p_2}{(2\pi)^3}
		\kk\cdot\LL_2(\pp_1,\pp_2) 
		\kk\cdot\LL_2(-\pp_1,-\pp_2) P_{\rm lin}(p_1) P_{\rm lin}(p_2) \nonumber \\
		&=& \left( \frac{3}{7} \right)^2\int \frac{d^3p_1}{(2\pi)^3}\int \frac{d^3p_2}{(2\pi)^3}
		\frac{(\kk\cdot\pp_{[1,2]})^2}{|\pp_{[1,2]}|^4}
		\left(  1 - \frac{(\pp_1\cdot\pp_2)^2}{p_1^2p_2^2} \right)^2 P_{\rm lin}(p_1)P_{\rm lin}(p_2) \nonumber \\
		&=& \frac{1}{3}\left( \frac{3}{7} \right)^2\int \frac{d^3p_1}{(2\pi)^3}\int \frac{d^3p_2}{(2\pi)^3}
		\frac{k^2}{|\pp_{[1,2]}|^2}
		\left(  1 - \frac{(\pp_1\cdot\pp_2)^2}{p_1^2p_2^2} \right)^2 P_{\rm lin}(p_1)P_{\rm lin}(p_2) \nonumber \\
       &=& \frac{3k^2}{392\cdot32\pi^4} \int \frac{dp_1}{p_1^3} \frac{dp_2}{p_2^3}
	  K(p_1,p_2) P_{\rm lin}(p_1) P_{\rm lin}(p_2), \nonumber \\
\end{eqnarray}
\begin{eqnarray}
		k^2\sigma_{\rm v,13}^2 &\equiv&
		-2 \int \frac{d^3p_1}{(2\pi)^3} \int \frac{d^3p_2}{(2\pi)^3}
		\kk\cdot \LL_1(\pp_1) 
		\kk\cdot \LL_3(-\pp_1,\pp_2,-\pp_2) P_{\rm lin}(p_1) P_{\rm lin}(p_2) \nonumber \\
		&=&	 \frac{10}{63} \int \frac{d^3p_1}{(2\pi)^3} \int \frac{d^3p_2}{(2\pi)^3}
		\frac{k^2}{p_1^2}\left( 1 - \frac{(\pp_1\cdot\pp_2)^2}{p_1^2p_2^2} \right)
		\left(   \frac{p_1^2p_2^2 - (\pp_1\cdot\pp_2)^2}{|\pp_1-\pp_2|^2 p_2^2} 
		+\frac{p_1^2p_2^2 - (\pp_1\cdot\pp_2)^2}{|\pp_1+\pp_2|^2 p_2^2}
		\right)P_{\rm lin}(p_1) P_{\rm lin}(p_2)\nonumber \\
		&=&	\frac{20}{63}\int \frac{d^3p_1}{(2\pi)^3} \int \frac{d^3p_2}{(2\pi)^3}
		\frac{k^2}{|\pp_1+\pp_2|^2}\left( 1 - \frac{(\pp_1\cdot\pp_2)^2}{p_1^2p_2^2} \right)^2P_{\rm lin}(p_1) P_{\rm lin}(p_2) \nonumber \\
       &=& \frac{5k^2}{126\cdot32\pi^4} \int \frac{dp_1}{p_1^3} \frac{dp_2}{p_2^3}
	    K(p_1,p_2)P_{\rm lin}(p_1) P_{\rm lin}(p_2), 
\end{eqnarray}
where
\begin{eqnarray}
		K(p_1,p_2) =\left( p_1^2-p_2^2 \right)^4\ln \left( \frac{(p_1+p_2)^2}{(p_1-p_2)^2} \right)
		-\frac{4}{3}p_1p_2(3p_1^6 - 11p_2^2p_1^4 - 11 p_1^2 p_2^4 + 3p_2^6).
\end{eqnarray}
The 1-loop velocity dispersion is given by $\sigma_{\rm v,1\mathchar`-loop}^2 \equiv \sigma_{\rm v,22}^2 + \sigma_{\rm v,13}^2$.

\section{Smoothed high-$k$ solutions}
\label{high-k}

As shown in eq.~\eqref{ap_P1loop},
the high-$k$ solution for $P_{22}$ has the BAO behavior owing to the linear power spectrum.
However, the exact solution for $P_{22}$ does not have the oscillation behavior because
it is computed by integrating the linear power spectrum.
To modify this incorrect BAO behavior for $P_{22}$,
we simply replace the linear power spectrum in the high-$k$ solution for $P_{22}$
with the no-wiggle linear power spectrum as follows:
\begin{eqnarray}
		P_{22}(k) \to \left( \frac{k^2 \sigma_{\rm v,lin}^2}{2} \right)P_{\rm lin}^{\rm nw}(k).
		\label{each_ap_1loop_nw}
\end{eqnarray}
This solution indeed behaves as the approximated solution of the exact $P_{22}$ 
at least than the high-$k$ solution for $P_{22}$ (see figure~\ref{fig:SPT2loop}),
and give the following approximated 1-loop solution
\begin{eqnarray}
		P_{\rm 1\mathchar`-loop}(k) \to 
		-\left( \frac{k^2 \sigma_{\rm v,lin}^2}{2} \right)\left( P_{\rm lin}(k) - P_{\rm lin}^{\rm nw}(k) \right).
\end{eqnarray}

Similarly, we have the following no-wiggle high-$k$ solution for $P_{33b}$ and $P_{24}$,
\begin{eqnarray}
	P_{33b}(k)		&\to&   \left( \frac{k^2 \sigma_{\rm v,lin}^2}{2} \right)P_{22}(k)
	-\frac{1}{2} \left(\frac{k^2 \sigma_{\rm v,lin}^2}{2}\right)^2 P_{\rm lin}^{\rm nw}(k)
	+ \left( \frac{k^2\sigma_{\rm v, 22}^2}{2} \right) P_{\rm lin}^{\rm nw}(k), \nonumber \\
	P_{24}(k) &\to& - \left( \frac{k^2\sigma_{\rm v,lin}^2}{2} \right)P_{22}(k),
	\label{each_ap_2loop_nw_1loop}
\end{eqnarray}
and these modified solutions have a more reasonable behavior as the approximation of the exact $P_{24}$ and $P_{33b}$
than the high-$k$ solutions (see figure~\ref{fig:SPT2loop}).
When we use $P_{22} \to k^2 \sigma_{\rm v,lin}^2 P_{\rm lin}^{\rm nw}/2$ and $P_{13} \to -k^2 \sigma_{\rm v,lin}^2 P_{\rm lin}/2$,
we have
\begin{eqnarray}
		P_{33b}(k) &\to& \frac{1}{2}\left( \frac{k^2 \sigma_{\rm v,lin}^2}{2} \right)^2 P_{\rm lin}^{\rm nw}(k)
		+ \left( \frac{k^2 \sigma_{\rm v,22}^2}{2} \right)P_{\rm lin}^{\rm nw}(k), \nonumber \\
		P_{24}(k) &\to& -\left( \frac{k^2 \sigma_{\rm v,lin}^2}{2} \right)^2 P_{\rm lin}^{\rm nw}(k) 
		+ \left(\frac{k^2\sigma_{\rm v,13}^2}{2}  \right) P_{\rm lin}^{\rm nw}(k).
		\label{each_ap_2loop_nw_linear}
\end{eqnarray}
Thus, these smoothed high-$k$ solutions and the high-$k$ solutions for $P_{15}$ and $P_{33a}$
give the approximated 2-loop solution as follows:
\begin{eqnarray}
		P_{\rm 2\mathchar`-loop}(k) \to \left( \frac{1}{2}\left( \frac{k^2 \sigma_{\rm v,lin}^2}{2} \right)^2
		- \left( \frac{k^2 \sigma_{\rm v,1\mathchar`-loop}^2}{2} \right)\right)\left( P_{\rm lin}(k) - P_{\rm lin}^{\rm nw}(k) \right).
\end{eqnarray}

\bibliographystyle{JHEP}

\providecommand{\href}[2]{#2}\begingroup\raggedright\endgroup
\end{document}